\documentstyle[12pt]{article}
\newread\epsffilein    
\newif\ifepsffileok    
\newif\ifepsfbbfound   
\newif\ifepsfverbose   
\newdimen\epsfxsize    
\newdimen\epsfysize    
\newdimen\epsftsize    
\newdimen\epsfrsize    
\newdimen\epsftmp      
\newdimen\pspoints     
\def\epsfbox#1{\global\def\epsfllx{72}\global\def\epsflly{72}%
   \global\def\epsfurx{540}\global\def\epsfury{720}%
   \def\lbracket{[}\def\testit{#1}\ifx\testit\lbracket
   \let\next=\epsfgetlitbb\else\let\next=\epsfnormal\fi\next{#1}}%
\def\epsfgetlitbb#1#2 #3 #4 #5]#6{\epsfgrab #2 #3 #4 #5 .\\%
   \epsfsetgraph{#6}}%
\def\epsfnormal#1{\epsfgetbb{#1}\epsfsetgraph{#1}}%
\def\epsfgetbb#1{%
\openin\epsffilein=#1
\ifeof\epsffilein\errmessage{I couldn't open #1, will ignore it}\else
   {\epsffileoktrue \chardef\other=12
    \def\do##1{\catcode`##1=\other}\dospecials \catcode`\ =10
    \loop
       \read\epsffilein to \epsffileline
       \ifeof\epsffilein\epsffileokfalse\else
          \expandafter\epsfaux\epsffileline:. \\%
       \fi
   \ifepsffileok\repeat
   \ifepsfbbfound\else
    \ifepsfverbose\message{No bounding box comment in #1; using defaults}\fi\fi
   }\closein\epsffilein\fi}%
\def\epsfsetgraph#1{%
   \epsfrsize=\epsfury\pspoints
   \advance\epsfrsize by-\epsflly\pspoints
   \epsftsize=\epsfurx\pspoints
   \advance\epsftsize by-\epsfllx\pspoints
   \epsfxsize\epsfsize\epsftsize\epsfrsize
   \ifnum\epsfxsize=0 \ifnum\epsfysize=0
      \epsfxsize=\epsftsize \epsfysize=\epsfrsize
     \else\epsftmp=\epsftsize \divide\epsftmp\epsfrsize
       \epsfxsize=\epsfysize \multiply\epsfxsize\epsftmp
       \multiply\epsftmp\epsfrsize \advance\epsftsize-\epsftmp
       \epsftmp=\epsfysize
       \loop \advance\epsftsize\epsftsize \divide\epsftmp 2
       \ifnum\epsftmp>0
          \ifnum\epsftsize<\epsfrsize\else
             \advance\epsftsize-\epsfrsize \advance\epsfxsize\epsftmp \fi
       \repeat
     \fi
   \else\epsftmp=\epsfrsize \divide\epsftmp\epsftsize
     \epsfysize=\epsfxsize \multiply\epsfysize\epsftmp   
     \multiply\epsftmp\epsftsize \advance\epsfrsize-\epsftmp
     \epsftmp=\epsfxsize
     \loop \advance\epsfrsize\epsfrsize \divide\epsftmp 2
     \ifnum\epsftmp>0
        \ifnum\epsfrsize<\epsftsize\else
           \advance\epsfrsize-\epsftsize \advance\epsfysize\epsftmp \fi
     \repeat     
   \fi
   \ifepsfverbose\message{#1: width=\the\epsfxsize, height=\the\epsfysize}\fi
   \epsftmp=10\epsfxsize \divide\epsftmp\pspoints
   \vbox to\epsfysize{\vfil\hbox to\epsfxsize{%
      \includegraphics{#1}%
      \hfil}}%
\epsfxsize=0pt\epsfysize=0pt}%

{\catcode`\%=12 \global\let\epsfpercent=
\long\def\epsfaux#1#2:#3\\{\ifx#1\epsfpercent
   \def\testit{#2}\ifx\testit\epsfbblit
      \epsfgrab #3 . . . \\%
      \epsffileokfalse
      \global\epsfbbfoundtrue
   \fi\else\ifx#1\par\else\epsffileokfalse\fi\fi}%
\def\epsfgrab #1 #2 #3 #4 #5\\{%
   \global\def\epsfllx{#1}\ifx\epsfllx\empty
      \epsfgrab #2 #3 #4 #5 .\\\else
   \global\def\epsflly{#2}%
   \global\def\epsfurx{#3}\global\def\epsfury{#4}\fi}%
\def\epsfsize#1#2{\epsfxsize}
\let\epsffile=\epsfbox

\def\ltap{\raisebox{-.4ex}{\rlap{$\sim$}} \raisebox{.4ex}{$<$}}

\newlength{\dinwidth}
\newlength{\dinmargin}
\setlength{\dinwidth}{21.0cm}
\textheight24.2cm \textwidth17.0cm
\setlength{\dinmargin}{\dinwidth}
\addtolength{\dinmargin}{-\textwidth}
\setlength{\dinmargin}{0.5\dinmargin}
\oddsidemargin -1.0in
\addtolength{\oddsidemargin}{\dinmargin}
 
\setlength{\evensidemargin}{\oddsidemargin}
\setlength{\marginparwidth}{0.9\dinmargin}
\marginparsep 8pt \marginparpush 5pt
\topmargin -42pt
\headheight 12pt
\headsep 30pt 
\parskip 3mm plus 2mm minus 2mm
\parindent 5mm
\renewcommand{\arraystretch}{1.3}
\renewcommand{\thefootnote}{\arabic{footnote}}

\newcommand{\bfig}{\begin{figure}}
\newcommand{\efig}{\end{figure}}
\newcommand{\bcen}{\begin{center}}
\newcommand{\ecen}{\end{center}}
\newcommand{\beq}{\begin{equation}}
\newcommand{\eeq}{\end{equation}}
\newcommand{\btabu}{\begin{tabular}}
\newcommand{\etabu}{\end{tabular}}
\newcommand{\btabl}{\begin{table}}
\newcommand{\etabl}{\end{table}}

\def\lsim{\mathrel{\rlap{\lower4pt\hbox{\hskip1pt$\sim$}}
    \raise1pt\hbox{$<$}}}         
\def\gsim{\mathrel{\rlap{\lower4pt\hbox{\hskip1pt$\sim$}}
    \raise1pt\hbox{$>$}}}         

\pagestyle{empty}
 
\begin{document}

\title{Study of Elastic $\rho^0$ Photoproduction at HERA\\
       using the ZEUS Leading Proton Spectrometer}
\author{ZEUS Collaboration
       }
\date{}
\markright{.}

\maketitle  
\pagestyle{empty}

\vspace{5 cm}
\begin{abstract}
The differential cross section $d\sigma/dt$ for elastic $\rho^0$ 
photoproduction, 
$\gamma p \rightarrow  \rho^0 p~(\rho^0 \rightarrow \pi^+ \pi^-)$,
has been measured in $ep$ interactions at HERA.
The squared four-momentum exchanged at the proton vertex, $t$,
has been determined directly by 
measuring the momentum of the scattered proton using the ZEUS 
Leading Proton Spectrometer (LPS), a large scale 
system of silicon micro-strip detectors operating 
close to the HERA proton beam.
The LPS allows the measurement of the momentum of high energy 
protons scattered at small angles with 
accuracies of 0.4\% for the longitudinal momentum and 
5~MeV for the transverse momentum.
Photoproduction of $\rho^0$ mesons has been 
investigated in the interval $0.073<|t|<0.40$~GeV$^2$, for 
photon virtualities 
$Q^2<1$~GeV$^2$ and photon-proton centre-of-mass energies $W$ between 
50 and 100~GeV. 
In the measured range, the $t$ distribution exhibits an exponential shape 
with a slope parameter $b = 9.8\pm 0.8~(\mbox{stat.}) 
\pm 1.1~(\mbox{syst.})~\mbox{GeV}^{-2}$. 
The use of the LPS eliminates 
the contamination from events with diffractive dissociation 
of the proton into low mass states.

\vspace{-20 cm}
{\noindent
DESY 96-183 \\
August 1996 }

\end{abstract}
\thispagestyle{empty}

\newpage

%
%
%
%
\topmargin-1.cm                                                                                    
\evensidemargin-0.3cm                                                                              
\oddsidemargin-0.3cm                                                                               
\textwidth 16.cm                                                                                   
\textheight 680pt                                                                                  
\parindent0.cm                                                                                     
\parskip0.3cm plus0.05cm minus0.05cm                                                               
\def\3{\ss}                                                                                        
\newcommand{\address}{ }                                                                           
\renewcommand{\author}{ }                                                                          
\pagenumbering{Roman}                                                                              
\pagestyle{plain}
                                                   %

\begin{center}                                                                                     
{                      \Large  The ZEUS Collaboration              }                               
\end{center}                                                                                       
  M.~Derrick,                                                                                      
  D.~Krakauer,                                                                                     
  S.~Magill,                                                                                       
  D.~Mikunas,                                                                                      
  B.~Musgrave,                                                                                     
  J.R.~Okrasi\'{n}ski,                                                                             
  J.~Repond,                                                                                       
  R.~Stanek,                                                                                       
  R.L.~Talaga,                                                                                     
  H.~Zhang  \\                                                                                     
 {\it Argonne National Laboratory, Argonne, IL, USA}~$^{p}$
% ...
\newpage                                                                                           
% ...
$^{\    2}$ now at Univ. of Salerno and INFN Napoli, Italy \\                                      
% ...
$^{  19}$ now at ATLAS Collaboration, Univ. of Munich \\                                           
% ...
                                                           %
                                                           %
\newpage   
                                                           %
                                                           %
\begin{tabular}[h]{rp{14cm}}                                                                       
$^{a}$ &  supported by the Natural Sciences and Engineering Research                               
          Council of Canada (NSERC)  \\                                                            
$^{b}$ &  supported by the FCAR of Qu\'ebec, Canada  \\                                            
$^{c}$ &  supported by the German Federal Ministry for Education and                               
          Science, Research and Technology (BMBF), under contract                                  
          numbers 057BN19P, 057FR19P, 057HH19P, 057HH29P, 057SI75I \\                              
$^{d}$ &  supported by the MINERVA Gesellschaft f\"ur Forschung GmbH,                              
          the Israel Academy of Science and the U.S.-Israel Binational                             
          Science Foundation \\                                                                    
$^{e}$ &  supported by the German Israeli Foundation, and                                          
          by the Israel Academy of Science  \\                                                     
$^{f}$ &  supported by the Italian National Institute for Nuclear Physics                          
          (INFN) \\                                                                                
$^{g}$ &  supported by the Japanese Ministry of Education, Science and                             
          Culture (the Monbusho) and its grants for Scientific Research \\                         
$^{h}$ &  supported by the Korean Ministry of Education and Korea Science                          
          and Engineering Foundation  \\                                                           
$^{i}$ &  supported by the Netherlands Foundation for Research on                                  
          Matter (FOM) \\                                                                          
$^{j}$ &  supported by the Polish State Committee for Scientific                                   
          Research, grants No.~115/E-343/SPUB/P03/109/95, 2P03B 244                                
          08p02, p03, p04 and p05, and the Foundation for Polish-German                            
          Collaboration (proj. No. 506/92) \\                                                      
$^{k}$ &  supported by the Polish State Committee for Scientific                                   
          Research (grant No. 2 P03B 083 08) and Foundation for                                    
          Polish-German Collaboration  \\                                                          
$^{l}$ &  partially supported by the German Federal Ministry for                                   
          Education and Science, Research and Technology (BMBF)  \\                                
$^{m}$ &  supported by the German Federal Ministry for Education and                               
          Science, Research and Technology (BMBF), and the Fund of                                 
          Fundamental Research of Russian Ministry of Science and                                  
          Education and by INTAS-Grant No. 93-63 \\                                                
$^{n}$ &  supported by the Spanish Ministry of Education                                           
          and Science through funds provided by CICYT \\                                           
$^{o}$ &  supported by the Particle Physics and                                                    
          Astronomy Research Council \\                                                            
$^{p}$ &  supported by the US Department of Energy \\                                              
$^{q}$ &  supported by the US National Science Foundation \\                                       
\end{tabular}                                                                                      
                                                           %
                                                           %

\newpage

\setlength{\dinwidth}{21.0cm}
\textheight24.2cm \textwidth17.0cm
\setlength{\dinmargin}{\dinwidth}
\addtolength{\dinmargin}{-\textwidth}
\setlength{\dinmargin}{0.5\dinmargin}
\oddsidemargin -1.0in
\addtolength{\oddsidemargin}{\dinmargin}
 
\setlength{\evensidemargin}{\oddsidemargin}
\setlength{\marginparwidth}{0.9\dinmargin}
\marginparsep 8pt \marginparpush 5pt
\topmargin -42pt
\headheight 12pt
\headsep 30pt 
\parskip 3mm plus 2mm minus 2mm
\parindent 5mm
\renewcommand{\arraystretch}{1.3}
\renewcommand{\thefootnote}{\arabic{footnote}}
\pagenumbering{arabic} 
\setcounter{page}{1}
\section{Introduction}

The physics of diffractive scattering processes
has emerged as one of the most interesting 
topics of study in the early running of HERA. Up to now, 
the cross section for events
in which the virtual photon diffractively dissociates
into a vector meson or a generic state $X$
has been measured in the H1 and ZEUS experiments either by requiring a 
``large rapidity gap" between the proton beam direction and 
the most forward energy deposit
recorded in the detector or by subtracting the non-diffractive 
background in a statistical way~\cite{hera_diffractive}-\cite{H1_rhopsi_hiq2}.
Here we present the first cross section measurement at
HERA in which diffraction is tagged by the detection of 
a high energy scattered proton, thereby eliminating contamination by events with
dissociation of the proton.

The measurement of the proton was performed using the ZEUS 
Leading Proton Spectrometer (LPS), which detects protons scattered at
very small angles ($\ltap 1$~mrad). 
In this spectrometer, silicon micro-strip detectors are 
used in conjunction with the proton beam line elements to measure
the momentum of the scattered proton. The detectors are positioned 
as close as the $10\sigma$ envelope of the circulating proton 
beam (typically a few mm) by using the ``Roman pot" technique~\cite{romanpot}.
In the configuration used to 
collect the data presented here, the LPS consisted of a total of about
22,000 channels.

This paper concentrates on the exclusive process $\gamma p \rightarrow 
\rho^0 p$ in $ep$ interactions at small photon virtualities 
($Q^2 \approx 0$, the 
``photoproduction" region). This reaction 
is often called ``elastic", in reference to the vector meson dominance
model (VDM).
Elastic photoproduction of 
$\rho^0$ mesons has been 
investigated in fixed target experiments at photon-proton 
centre-of-mass energies $W \lsim 20$~GeV~\cite{bauer}-\cite{omega} as well
as at HERA energies, 
$W \approx 100$-200~GeV~\cite{rho93,rhoh1}.
The process has the characteristic features of 
soft diffractive interactions: the dependence of the cross section on $W$ 
is weak, the dependence on $t$ is approximately exponential, and the 
vector meson is observed to retain the helicity of the photon 
($s$-channel helicity conservation). Here $t$ is the squared 
four-momentum exchanged at the proton vertex.
The data presented in this paper cover the kinematic range 
$50<W<100$~GeV, $Q^2 < 1$~GeV$^2$ and $0.073<|t|<0.40$~GeV$^2$.
Elastic events were selected by requiring that the scattered proton 
carry more than 98\% of the incoming proton beam energy.
The scattered positron was not detected; 
however, $Q^2$ was estimated using transverse momentum balance. 


\section{Experimental set-up}
\label{setup}

\subsection{HERA}

The data discussed here were collected in 1994 at 
HERA which operated with 820~GeV protons and 27.5~GeV positrons (indicated in 
the following with the symbol $e$). The 
proton and positron beams each contained 153 colliding bunches, together with 
17 additional unpaired proton and 15 unpaired positron bunches. These 
additional bunches were used for background studies. The time between 
bunch crossings was 96~ns. The typical instantaneous luminosity was
$1.5 \times 10^{30}$~cm$^{-2}$s$^{-1}$ and the integrated luminosity for 
this study is $898\pm 14$~nb$^{-1}$.

\subsection{The ZEUS detector}

A detailed description of the ZEUS detector can be found 
elsewhere~\cite{detector_a}.
A brief outline of the components in the central ZEUS 
detector~\cite{detector_b} which are 
most relevant for this analysis is given below, followed by a description
of the Leading Proton Spectrometer.

\subsubsection{Central components and luminosity measurement}

Charged particles are
tracked by the inner tracking detectors which operate in a
magnetic field of 1.43 T provided by a thin superconducting coil.
Immediately surrounding the beam pipe is the vertex detector (VXD), 
a drift chamber which consists of 120 radial cells, each with 12 
sense wires~\cite{vxd}. 
It is surrounded by the central tracking detector (CTD), which consists 
of 72 cylindrical drift chamber layers,  organised into 9 
superlayers covering the polar angle region 
$15^\circ < \theta < 164^\circ$\footnote{The 
coordinate system used in this paper has the 
$Z$ axis pointing in
the proton beam direction, hereafter referred to as ``forward'',
the $X$ axis pointing horizontally towards the centre of HERA and
the $Y$ axis pointing upwards. The polar angle
$\theta$ is defined with respect to the  $Z$ direction.}~\cite{ctd}. 

The high resolution uranium-scintillator calorimeter (CAL) \cite{CAL}  
consists of three parts: the
forward (FCAL), the rear (RCAL) and the barrel calorimeter (BCAL).
Each part is subdivided transversely into towers and
longitudinally into one electromagnetic section (EMC) and one (in RCAL)
or two (in BCAL and FCAL) hadronic sections (HAC). A section of a tower
is called a cell; each cell is viewed by two photomultiplier tubes.
The CAL energy resolution, as measured under test beam conditions,
is $\sigma_E/E=0.18/\sqrt{E}$ for electrons and $\sigma_E/E=0.35/\sqrt{E}$ 
for hadrons ($E$ in GeV). 

The Veto Wall, the C5 counter and the small angle rear tracking detector (SRTD)
all consist of scintillation counters and
are located at $Z =- 730$~cm, $Z =- 315$~cm and $Z =- 150$~cm, 
respectively. 
Particles which are generated by proton beam-gas interactions upstream
of the nominal $ep$ interaction point hit the RCAL, the Veto Wall, the 
SRTD and C5 at different times 
than particles originating from the nominal $ep$ interaction point.
Proton beam-gas events are thus rejected by timing measurements
in these detectors.


The luminosity is determined from the rate of the Bethe-Heitler 
process, $ep \rightarrow e \gamma p$, where the photon is measured with a 
calorimeter (LUMI) located in the HERA tunnel downstream of the 
interaction point in the direction of the outgoing positrons~\cite{lumi}.

\subsubsection{The Leading Proton Spectrometer}

The Leading Proton Spectrometer~\cite{detector_a} (LPS) detects
charged particles scattered at small angles and carrying a substantial fraction,
$x_L$, of the incoming proton momentum; these particles remain in the beam pipe
and their trajectory is measured by a system of position sensitive silicon 
micro-strip detectors very close to the proton beam. 
The track deflection induced by the magnets in the 
proton beam line is used for the momentum analysis of the scattered proton.

\begin{figure}
\vspace{-1.5cm}
\begin{center}
\leavevmode
\hbox{%
\hspace*{-0.8cm}
\epsfxsize = 15cm
\epsffile{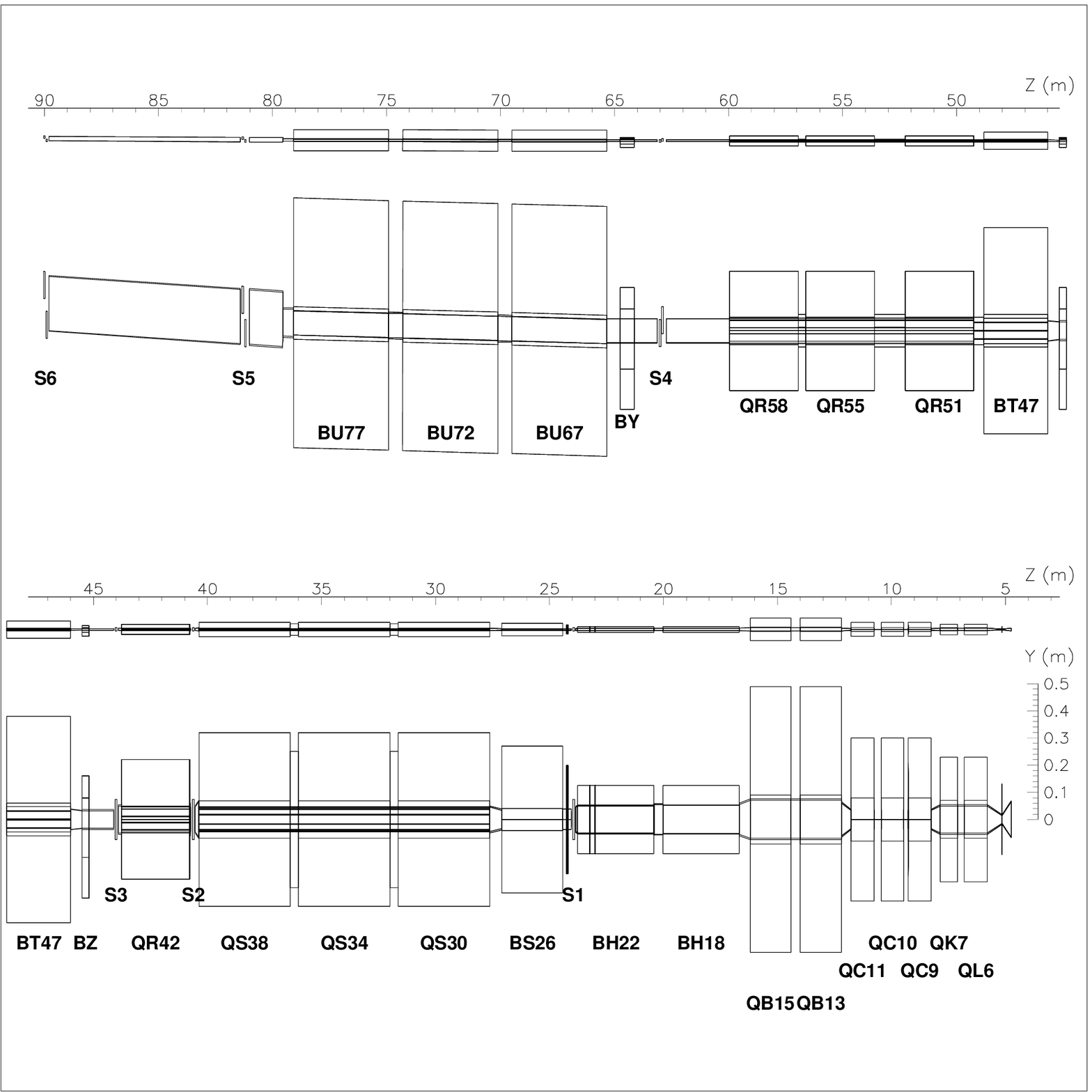}}
\end{center}
\vspace{-0.5cm}
\caption{Side view of the outgoing proton beam line for $Z$ between 
5 and 50~m (bottom) and between 50 and 90~m (top).
The small drawing below the $Z$ axis is to scale; in the large drawing 
different scales are used for the longitudinal and transverse 
directions. Magnetic elements labelled BH, BS  and BT are 
horizontally bending dipoles; vertically bending dipoles are indicated 
as BZ, BY and BU.
Quadrupoles are labelled as QL, QK, QC, QB, QS, QR;
the magnets upstream
of the septum magnet BS are common to the proton and positron beam
line. The positions of the LPS stations S1 through S6 are also shown. The
centre of the ZEUS detector is at $Z=0$.
}
\label{lps_detailed} 
\end{figure}

The layout of the LPS is shown in Fig.~\ref{lps_detailed}; it 
consists of six detector stations, 
S1 to S6, placed along the beam line in the direction of the outgoing 
protons, at $Z=23.8$~m, 40.3~m, 44.5~m, 
63.0~m, 81.2~m and 90.0~m from the interaction point, respectively.

\begin{figure}
\vspace{-1.0cm}
\begin{center}
\leavevmode
\hbox{%
\hspace*{-0.8cm}
\epsfxsize = 12cm
\epsffile{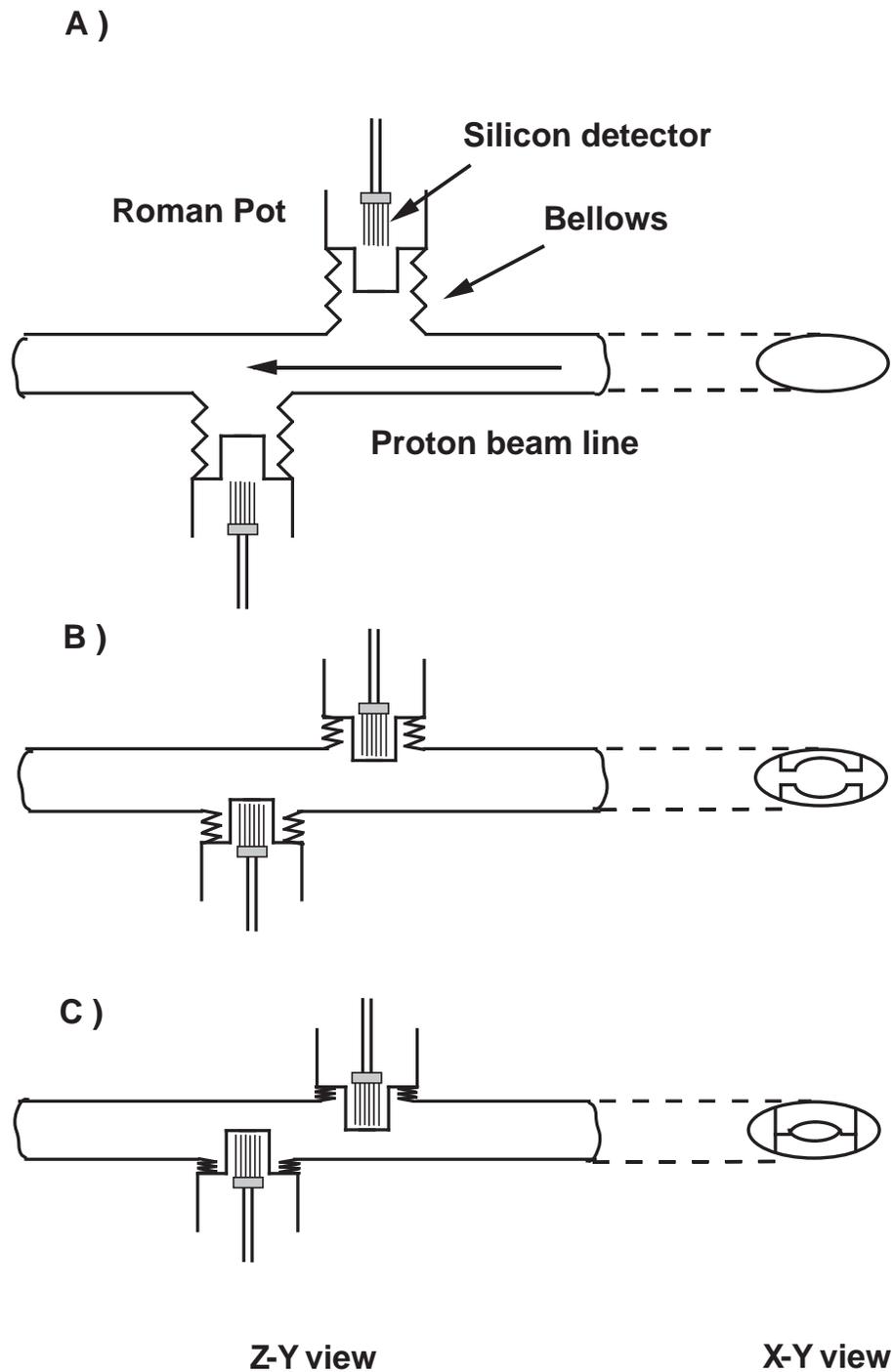}}
\end{center}
\vspace{-0.5cm}
\caption{Schematic layout of a station (like S4, S5 or S6). 
A) During beam filling and ramping, the detector planes (labelled
``Silicon detector") are kept outside of the pots and the pots are placed 
far from the beam. The zig-zag lines indicate the bellows.
B) The detector planes are inside the pots and the pots are being moved 
towards the beam. Note the elliptical profile of the fronts of the pots
($X$-$Y$ view), which 
matches the cutout of the detector planes. 
C) When taking data, the pots are fully inserted and the 
detector planes in the upper and lower half of the station 
partially overlap in the transverse plane. 
}
\label{pots}
\end{figure}

Each of the stations S1, S2 and S3 is equipped with an assembly of 
six planes of silicon micro-strip detectors parallel to each other
and mounted on a mobile arm, which allows them to be positioned near the proton 
beam. 
Stations S4, S5 and S6 each consist of two halves, 
each half containing an assembly of six planes similar to those of 
S1, S2, S3, also mounted on mobile arms, as shown in Fig.~\ref{pots}.
Each assembly has two planes with strips parallel
to the direction of motion of the arm,
two planes with strips at $+ 45^{\circ}$ and two at $- 45^{\circ}$ with respect 
to it; this makes it possible to measure the particle 
trajectory in three different projections in each assembly.
The dimensions of the detector planes vary from station to station 
but are approximately $4 \times 6$~cm$^2$.
The pitch is 115~$\mu$m for the planes with vertical or horizontal 
strips and
$115/\sqrt{2}=81~\mu$m  for the planes with $\pm 45^{\circ}$ strips.
The distance along $Z$ between neighbouring planes in an assembly is $\approx 7$~mm.
The detector planes are mounted in each assembly with a precision of
about $30$~$\mu$m.

The detector planes are inserted
in the beam pipe by means of re-entrant ``Roman pots" which allow the 
planes to operate at atmospheric pressure.  
A pot consists of a stainless steel cylinder with an open end away from the
beam; the other end is closed. The silicon detector planes are inserted
from the open end and are moved in until they are at about 0.5~mm from the 
closed end. The whole cylinder can be inserted transversely into the beam 
pipe. Figure~\ref{pots} illustrates the principle 
of operation.  The walls of the pots are  3~mm thick, except 
in front of and behind the detector planes, where they are 
400~$\mu$m thick; the thickness of the pot bottom walls facing the beam is
also 400~$\mu$m. The vacuum seal to the proton beam pipe 
is provided by steel bellows.
The pots and the detector planes are positioned by 
remotely controlled motors and are retracted during the
filling operations of the collider to increase the aperture of the 
vacuum chamber; this also minimises the radiation damage to the 
detectors and the front-end electronics. 
In stations S1, S2, S3 the detector planes 
are inserted into the beam pipe horizontally from the outside of the HERA 
ring towards the centre. In stations S4, S5, S6, the detector planes
in the two halves of each station independently approach the beam from above 
and from below. In the operating position the upper and lower 
halves partially overlap (cf. Fig.~\ref{pots}). The offset 
along the beam direction between the centres of the upper and lower pots 
is $\approx 10$~cm. Stations S5 and S6 were used in an earlier 
experiment at CERN and were adapted to the HERA beam line~\cite{ua4}.

Each detector plane has an elliptical cutout which 
follows the profile of the 
10$\sigma$ envelope of the beam, where~$\sigma$ is the standard deviation 
of the spatial distribution of the beam in the transverse plane.
Since the 10$\sigma$ profile differs from station to station, 
the shape of the cutout varies from station to station;
in data taking conditions the distance of each detector from the beam 
centre is also different and ranges from 3 to 20~mm. 
Small variations of the detector positions
from fill to fill are necessary during operation in order to follow the 
changes of the beam position 
and adapt to the background conditions.

The detector planes are read out by two types of VLSI chips mounted on the 
detector support:
a bipolar amplifier-comparator~\cite{tekz} followed by a radiation hard 
CMOS digital pipeline~\cite{dtsc}, which operates with a clock 
frequency of 10.4~MHz, synchronous with the HERA bunch crossing. Each 
chip has 64 channels reading out 64 adjacent strips.
The chips are radiation hard up to doses of several
Mrad.

\bigskip

\begin{figure}
\vspace{-1.0cm}
\begin{center}
\leavevmode
\hbox{%
\hspace*{-0.8cm}
\epsfxsize = 12cm
\epsffile{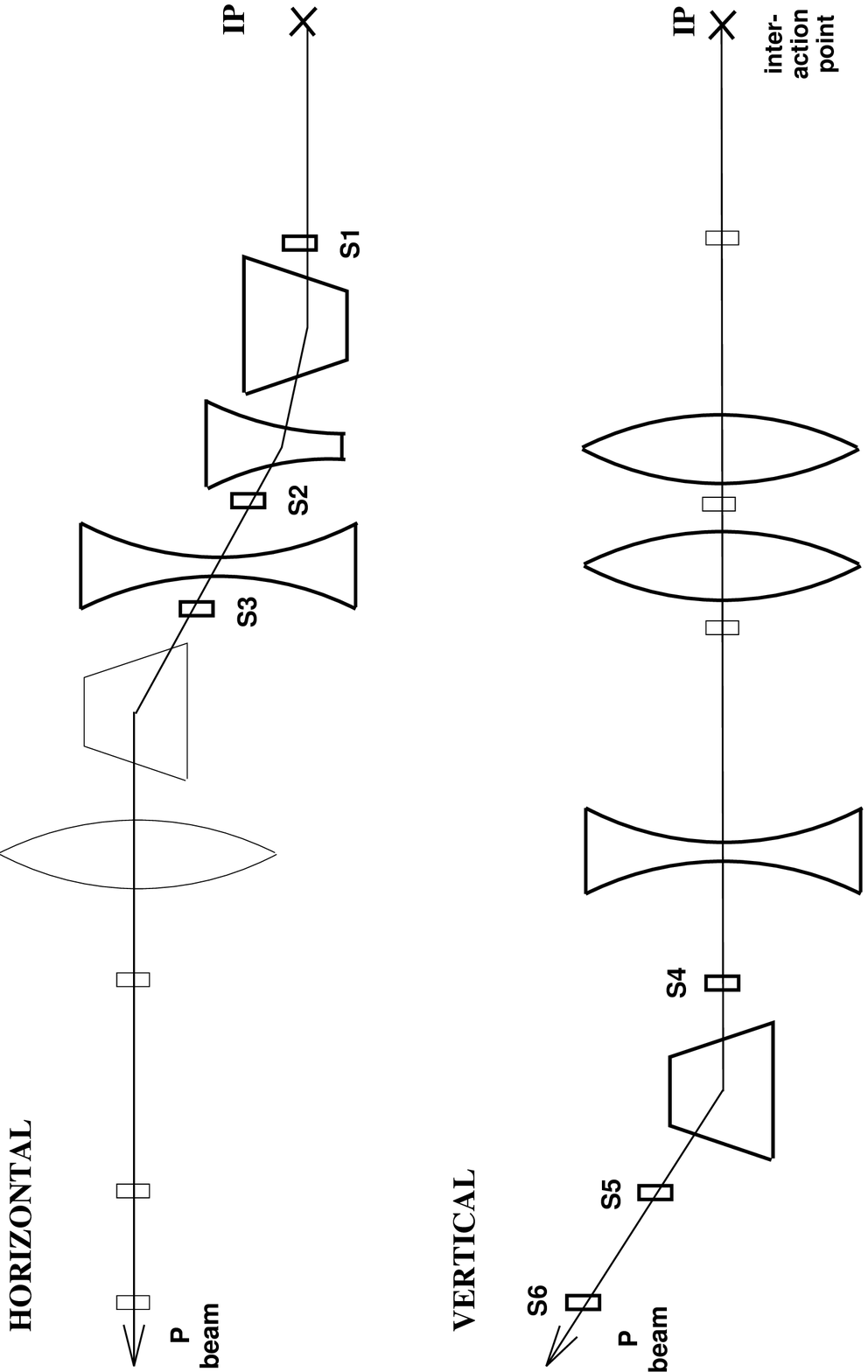}}
\end{center}
\vspace{-0.5cm}
\caption{Simplified diagram of the HERA proton beam optics relevant 
for the LPS. The top view (labelled ``horizontal") shows the magnetic elements 
relevant to the S1, S2 and S3 spectrometer as thick lines. Note
the half quadrupole between S1 and S2. The side view
(labelled ``vertical") shows the magnetic elements 
relevant to the S4, S5 and S6 spectrometer as thick lines.
}
\label{lps}
\end{figure}

A simplified 
diagram of the spectrometer optics is shown in Fig.~\ref{lps},
in which the beam line elements 
have been combined to show the main 
optical functions. Together with the HERA proton beam magnets, 
the six LPS stations form two spectrometers: 
\begin{enumerate}
\item Stations S1, S2, S3 use the combined horizontal bending power of a 
septum magnet and three magnetic septum half-quadrupoles.  
S1, S2, S3 were not operational in 1994 and are not discussed further here.

\item Stations S4, S5, S6 exploit in addition the vertical bending provided by 
three main dipole magnets (BU). These stations were used for the present 
measurement. 

\end{enumerate}

The insertion of the detectors into the operating positions 
typically begins as soon as the beams are brought into collision. Among 
the conditions required prior to beginning the insertion are the 
following: (i) proton beam position as measured with the HERA beam position 
monitor next to S4 within  1~mm of the nominal position;
(ii) background levels as measured in counters downstream of the main proton 
beam collimators, in the C5 counter and in the trigger counters of the 
Forward Neutron Calorimeter~\cite{FNC} (located downstream of 
S6 at $Z \approx 109$~m) stable and below 
given thresholds. About fifty minutes were necessary in 1994 
to insert the detector planes. This and the fact that the beam 
conditions did not always allow safe insertion of the detectors
results in the reduced value of the integrated luminosity available 
for this analysis with respect to other analyses of the ZEUS 1994 data.

The strip occupancy during data taking, i.e. the average number of 
strips firing per trigger divided by the total number of strips,
depended on the beam conditions but was 
typically less than 0.1\%, with small contributions from noise 
and synchrotron radiation.   

The fraction of noisy and malfunctioning 
channels in 1994 was less than 2\%; they were due to bad detector strips and
dead or noisy front-end channels. The efficiency of the detector planes,
after excluding these channels, was better than 99.8\%.

The LPS accepts
scattered protons carrying a fraction of the beam momentum, $x_L$, 
in the range $x_L \gsim 0.4$ and with $0 \lsim p_T \lsim 1$~GeV, 
where $p_T$ is the transverse momentum of the proton with respect to 
the incoming beam direction.
With the configuration installed in 1994 (S4, S5, S6), the 
resolution in $x_L$ is better than 
0.4\% at 820~GeV and the $p_T$ resolution is about 5~MeV. The latter is
less than the intrinsic transverse momentum spread in the
proton beam at the interaction point 
(with rms of about 40~MeV horizontally and about $90$~MeV 
vertically) due to the beam divergence of $\approx 50$~$\mu$rad 
in the horizontal and $\approx 110$~$\mu$rad in the vertical plane.
The LPS resolution is further discussed in section~\ref{offline}.

\subsubsection{Reconstruction of an LPS track}
\label{reconstruction}

\begin{figure}
\vspace{-1.0cm}
\begin{center}
\leavevmode
\hbox{%
\hspace*{-0.8cm}
\epsfxsize = 15cm
\epsffile{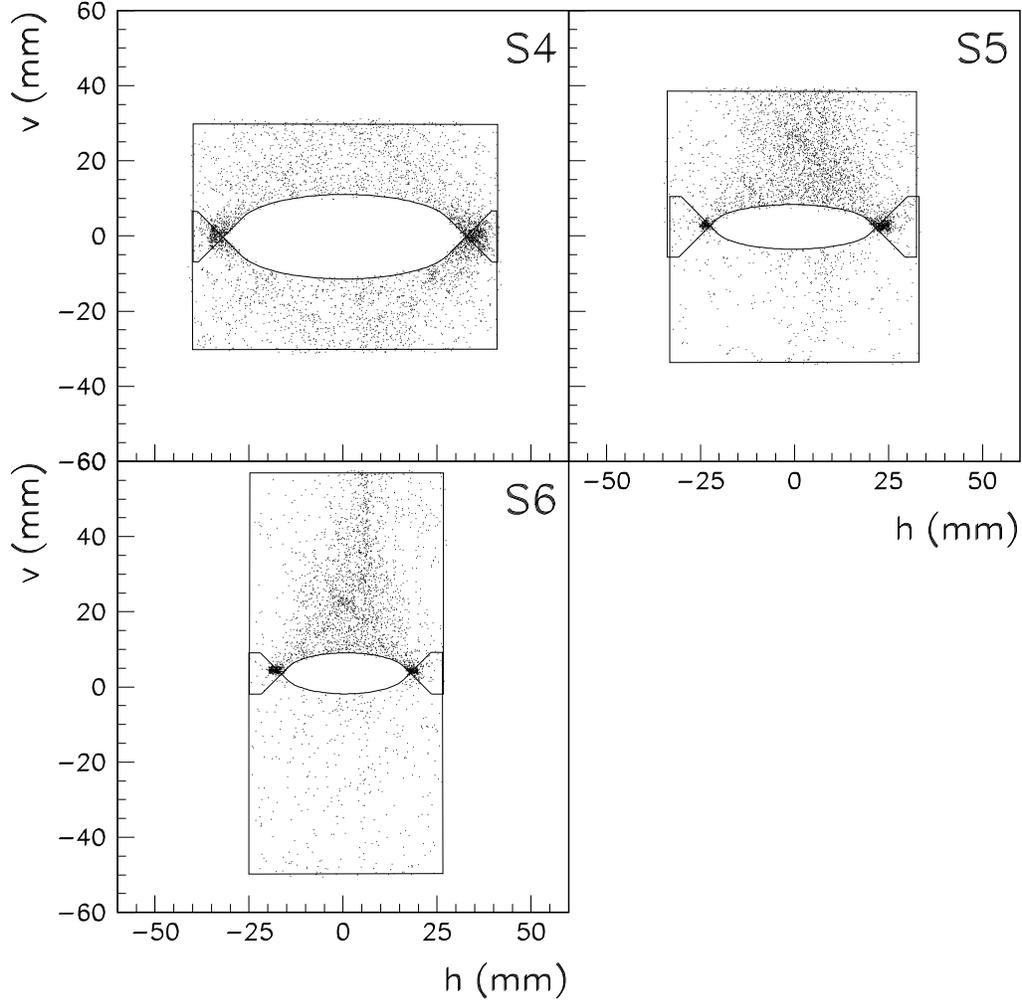}}
\end{center}
\vspace{-1cm}
\caption{Positions of the reconstructed track impact points 
in S4, S5 and S6. For each plot the origin of the reference frame coincides with 
the position of the nominal proton beam at the value of $Z$ 
corresponding to the centre of the station.
The continuous lines approximately indicate the sensitive region 
of the detector planes. 
}
\label{tesi3_8}
\end{figure}

Tracks are reconstructed in stages, proceeding from individual hits 
to full tracks~\cite{tesi_roberto}.
Noisy and malfunctioning channels are masked out and clusters of
adjacent hit strips are searched for in each detector plane.
Most clusters are one strip wide only (typically $\approx 25\%$ of the
clusters have more than 1 strip).
Track segments are then found independently in each detector assembly
of six planes.
As a first step, matching clusters in the two planes with
the same strip orientation are combined.
Candidate local track segments are then found by combining pairs of clusters
belonging to different projections; when a pair of such
clusters intersects within the region covered by the sensitive area of 
the detectors, a corresponding cluster in the remaining projection is 
searched for.
In order to reduce the number of candidates, local track segments that
traverse the overlap region of the detectors in the upper and the 
lower halves of the station are treated 
as one candidate.  Finally, all hits belonging to a candidate (up to twelve 
for tracks crossing the two halves, up to six otherwise)
are used in a fit to find the transverse coordinates
of the track at the value of $Z$ corresponding to the centre of the 
station. The spatial resolution on these coordinates is about
30~$\mu$m. 
Figure~\ref{tesi3_8} shows the position of the 
reconstructed coordinates in the stations S4, S5 and S6 for a 
typical run. The regions with a high density of reconstructed hits 
in the overlap zone between the upper and the lower detectors 
correspond to tracks with $x_L$ close to unity. Lower $x_L$ tracks 
are deflected upwards and focussed horizontally onto a vertical line.
For $x_L$ close to unity, this focus line is downstream of S6;
it approaches S6 as $x_L$ decreases and reaches S6 for $x_L\approx 0.7$.
This explains the fact that for low $x_L$ tracks, the impact points
in S5 and S6 tend to lie in a region which becomes narrower as the 
vertical coordinate increases.

We distinguish two classes of events: those which are detected in all three of 
the stations and those which are detected in only two stations.
In the latter, the interaction vertex 
position is required as a third point to measure the momentum. Tracks 
detected in three stations can be extrapolated backwards to 
$Z=0$ to also measure the transverse position of the interaction vertex.
In both cases, coordinates reconstructed in pairs of different stations are
first combined into track candidates and the track momentum is determined using
the average $ep$ interaction 
vertex with coordinates $(X_0,Y_0)$, found on a run-by-run basis with 
the sample of three-station tracks.  
Linear matrix equations relate the horizontal and vertical coordinates 
of the positions ($h_k,v_k$) and slopes ($h^{\prime}_k=dh_k/dl$,
$v^{\prime}_k=dv_k/dl$) of the track at each station to the position 
($X_0,Y_0$) and slope ($X_0^{\prime},Y_0^{\prime}$) of the track at the 
interaction point. The coordinate along the 
beam trajectory is $l$. The positions $(h_k,v_k)$ and slopes 
($h_k^{\prime},v_k^{\prime}$) are relative to the nominal beam position and 
direction at that value of $l$. 
The nominal beam crosses the interaction 
point ($Z=0$) at $X=Y=0$. For the horizontal direction one has:

\begin{eqnarray}
\left( \begin{array}{c}
h_k \\
h^{\prime}_k
\end{array}\right)=
\left( \begin{array}{cc}
m_0 & m_1 \\
m_2 & m_3
\end{array}\right) 
\left( \begin{array}{c}
X_0 \\
X^{\prime}_0
\end{array}\right)+
\left( \begin{array}{c}
b_0 \\
b_1
\end{array}\right).
\label{matrix}
\end{eqnarray}
 
\noindent 
An independent equation of the same form 
can be written for the vertical direction.
The matrix elements 
$m_{i}$ are known functions of $x_L$
which describe the beam optics including the effects 
of quadrupoles and drift 
lengths. The quantities $(b_0,b_1)$, also functions of
$x_L$, describe the deflection induced by the dipoles and by the 
quadrupoles in which the beam is off axis; since the beam is taken as reference,
they vanish as $x_L \rightarrow 1$.

Equation~(\ref{matrix}) and the corresponding one for the vertical 
direction can be written for a pair of stations $(a,b)$; 
upon eliminating the unknowns $X^{\prime}_0$ and $Y^{\prime}_0$, one finds

\begin{eqnarray}
h_b &= &M^{ab}_h(x_L) h_a + C^{ab}_h(x_L,X_0),\label{matrix01}\\
v_b &= &M^{ab}_v(x_L) v_a + C^{ab}_v(x_L,Y_0),
\label{matrix02}
\end{eqnarray}

\noindent
where $M^{ab}$ and $C^{ab}$ are functions of the matrix elements
$m_i$ and $b_i$.
These two equations are independent, apart from the common
dependence on $x_L$, and can be used to obtain two 
independent estimates of $x_L$.
If the values obtained are compatible, the pair of coordinates
is retained as a candidate two-station track.

As a final step
of the pattern recognition, three-station track candidates are 
searched for using pairs of two-station candidates, e.g. one in S4, S5 and
another in S5, S6. If a pair uses the same hits in the station 
common to the two tracks, 
if the projections of the two tracks on the horizontal (non-bending) plane 
coincide and 
if the momenta assigned to each track are compatible, then the 
two candidates are merged in a three-station track candidate.
Two-station and three-station candidates are then 
passed to a conventional track-fitting stage. 

A track $\chi^2$ is
defined as
\begin{equation}
\chi^2 = \left[ \sum_{i}
     {{(s_i - S_i(\psi))^2} \over {\sigma_i^2}}\right] 
          + {{(X_V - X_0)^2} \over {\sigma_{X_V}^2}}
               + {{(Y_V - Y_0)^2} \over {\sigma_{Y_V}^2}},
\label{eq:chisquared}
\end{equation}
\noindent
where the sum runs over all clusters in all planes assigned to a track. Here 
$s_i$ is the cluster position, $\sigma_i$ the uncertainty associated to it
(which includes the effects of multiple scattering and the 
contribution of the cluster width; typical values range from 50 to
100~$\mu$m),
$(X_V,Y_V)$ the interaction vertex coordinates in the $X,Y$ plane, 
$\sigma_{X_V}$ and $\sigma_{Y_V}$ the nominal widths of the 
vertex distribution;
$S_i$, a function of the five track parameters
${\bf{\psi}}=(X_V,Y_V,X_V^\prime,Y_V^\prime,x_L)$, is the predicted cluster
position calculated from equation~(\ref{matrix}) and the corresponding one
in the vertical direction; the quantities $X_V^\prime$, $Y_V^\prime$
indicate the track slopes at the interaction vertex.
The last two terms in eq.~(\ref{eq:chisquared}) constrain the track
to the interaction vertex.
This $\chi^2$ is minimised with respect to the five track parameters,
and the best track parameters, together with the error matrix, are determined.
In the present analysis, for $x_L$ close to unity, 
the average value of $\chi^2/ndf$ is
$\approx 1$, where $ndf$ is on average 7.3 for two-station tracks and 17.3
for three-station tracks. Three-station tracks are 60\% of the total.



\subsubsection{Alignment of the LPS}
\label{alignment}

The alignment of the LPS relies on survey 
information for locating the detector planes in $l$ and on high-energy proton 
tracks for locating them in $h$ and $v$.  
The individual detector planes are first aligned within one station,
then the relative alignment of the stations 
is determined, and finally
the three stations S4, S5, S6 are aligned relative to the ZEUS 
detector.  Typical accuracies in $h$ and $v$ are better than 20~$\mu$m. 
The actual path of the proton beam is also determined. These steps are
described below.

Tracks traversing the region in which the active areas of the detector 
planes in the upper and lower halves of a station overlap are used 
to align the detector planes within each half as well as to determine 
the position of the upper with respect to the lower half. With this 
procedure each plane is aligned independently; rotations of the 
detectors around the $l$ axis are also determined.  

The relative alignment between the S4, S5, S6 stations in $h$ 
is then found by exploiting the 
fact that tracks are straight lines in this projection.

The only magnetic elements between S4 and S6 are the 
dipoles between S4 and S5 which deflect particles vertically.
A sample of tracks with known deflection (i.e. known momentum) 
is thus necessary to align the stations relative to each other in $v$. 
This can be obtained independently of the LPS using the ZEUS calorimeter:

\begin{eqnarray}
x_L^{CAL}=1-\sum_i(E_i+p_{Zi})/(2E_p),
\label{align_x_l}
\end{eqnarray}

\noindent
where the sum runs over all calorimeter cells, $E$ is the energy measured 
in each cell and $p_Z=E\cos{\theta}$, with $\theta$ the polar angle of
each cell. The symbol $E_p$ denotes the incoming proton energy. 
Equation~(\ref{align_x_l}) follows from energy and momentum conservation: 
$\sum(E+p_Z)_{IN}=\sum(E+p_Z)_{OUT}$, where the sums run 
over the initial and final state particles, respectively, and 
$\sum(E+p_Z)_{IN}=2E_p$. 
Events are selected with $x_L^{CAL}>0.99$; these events have a clear peak in the
$x_L$ spectrum as measured by the LPS, with very little background underneath.
The relative positions of the stations are adjusted so that the peak appears
at $x_L$ of unity.
For the 1994 data, about 20,000 events were used.
The vertical alignment is finally checked by using events with elastic 
photoproduction of $\rho^0$ mesons -- those 
discussed in the present paper -- exploiting the fact that these events
have scattered protons with $x_L$ very close to unity: $x_L$ can be written,
for elastic $\rho^0$ photoproduction, as
$x_L  = 1 - (Q^2 + M_{\rho}^2 + |t|)/W^2$, where 
$M_{\rho}$ is the $\rho^0$ meson mass; 
for the sample used, the value of $x_L$ differs from unity by at most 0.2\%.

In order to align S4, S5 and S6 with respect to the proton beam line, tracks 
traversing all three stations are extrapolated to $Z=0$, taking into account 
the fields of all traversed magnetic 
elements (mostly quadrupoles, as shown in Fig.~\ref{lps_detailed}).  The detectors are
aligned with respect to the quadrupole axes by requiring that,
independent of $x_L$,
the average position of the extrapolated vertex be
the same as that measured by the central tracking detectors.
At this point the detectors are aligned
relative to the proton beam line and to the HERA quadrupole axes, and 
hence to the other components of ZEUS. About 40,000 three-station tracks
were used for this procedure.

\begin{figure}
\vspace{-1.0cm}
\begin{center}
\leavevmode
\hbox{%
\hspace*{-0.8cm}
\epsfxsize = 18cm
\epsffile{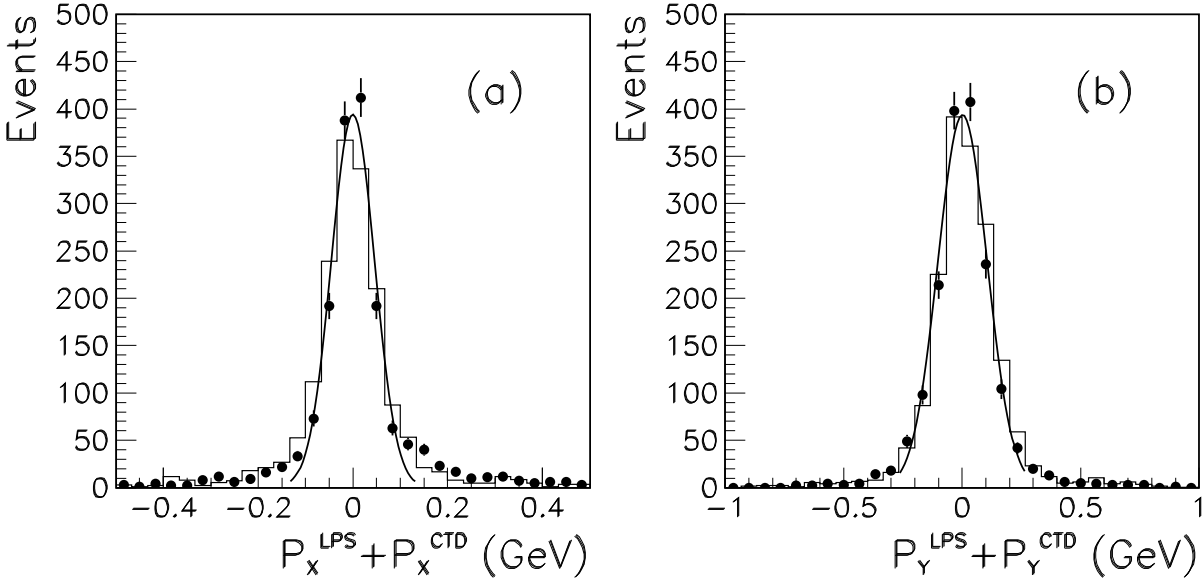}}
\end{center}
\vspace{-8cm}
\caption{Sum of the $X$ (a) and $Y$ (b) components of the proton
momentum as measured by the LPS and of the $\rho^0$ momentum as measured 
by the CTD.
The histogram was obtained with 
the Monte Carlo simulation discussed in section~\protect\ref{montecarlo}. The 
continuous line is the result of a Gaussian fit to the data. 
The fitted values
of the standard deviations are 45 and 102~MeV, respectively. They are 
dominated by the spread of the transverse momentum in the beam.
The other minor 
contributions are the LPS and CTD resolutions and the fact that $Q^2$ is 
not exactly zero; events with $Q^2\gsim 0.01$~GeV$^2$ contribute to the non-Gaussian tails.
}
\label{ctdlps}
\end{figure}

Finally, the average angle of the proton beam with respect 
to the nominal beam direction is determined by using events of elastic 
photoproduction of $\rho^0$ mesons. 
For such events the transverse components of the scattered proton 
momentum balance on average those of the $\rho^0$ meson.
The mean value of the sum of $p_X^{LPS}$ and $p_X^{CTD}$, and similarly
for $p_Y$,
is set to zero by adding a constant offset to the fitted angle of the
LPS tracks at the 
interaction vertex for all events. Here $p_X^{LPS}$ and $p_X^{CTD}$ indicate
the $X$ component of the proton momentum as measured by the LPS and 
of the $\rho^0$ momentum as measured by the CTD, respectively.
This procedure defines the direction of the $Z$ axis.
Typical values of the beam 
offset are $-15$~$\mu$rad and $-100$~$\mu$rad in the horizontal and 
vertical directions, respectively, with respect to the nominal
beam direction. The 1994 running period was split 
into three parts during which the beam tilt was relatively constant and
the offset was determined for each part. 
Fig.~\ref{ctdlps} shows, separately for the $X$ and $Y$ 
projections, the sum of the proton and the $\rho^0$ transverse momenta after 
this correction, which is determined by requiring that both histograms be
centred on zero. The width of the distributions is dominated by the 
intrinsic
spread of the transverse momentum in the beam. The other (minor) 
contributions are the LPS and CTD resolutions and the fact 
that the transverse momentum of the scattered positron is not 
identically zero since $Q^2$ is not zero. Note that the effect of 
non-zero $Q^2$ is just to widen the
distributions of Fig.~\ref{ctdlps}, not to shift them, 
since the $X$ and $Y$ components
of the scattered positron momentum are centred on zero. In addition
events with $Q^2\gsim 0.01$~GeV$^2$ contribute to the non-Gaussian tails.
The $x_L$ scale is not affected
by this tilt correction. The sensitivity of the determination of 
$t$ to the value of the tilt is weak, as discussed in section~\ref{results}. 
The effect of this correction is negligible for all quantities
measured in the central ZEUS apparatus.

As mentioned earlier, the detectors are in the retracted position 
between HERA runs; 
the positions of the pots (and hence of the detector planes) 
vary from one proton fill to the next by up to a few millimeters 
in $Y$ (rarely in $X$) depending on the beam position and on 
the background conditions. Coordinate reconstruction can thus 
not be more accurate than the reproducibility of the detector 
positioning system folded with the alignment accuracy.
This is monitored by the run-to-run dependence of the difference 
between the coordinates of the track 
impact point as measured by the 
detector planes in the upper and lower halves of a station 
for tracks in the overlap region. Note that this can be done since the 
alignment procedure described above is carried out using data from the
whole running period, i.e. not on a run-by-run basis.
The rms value of this difference is $\approx 25~\mu$m 
and is consistent 
with the specifications of the mechanics and commensurate with the detector 
resolution.

\section{Analysis}
\label{analysis}

\subsection{Event selection}
\label{event_selection}

\subsubsection{Trigger}

ZEUS uses a three-level trigger system~\cite{detector_a,detector_b}.
For the present data, the trigger 
selected events with photoproduction of a vector meson decaying 
into two charged particles with no requirement that either 
the scattered positron or the scattered proton be detected.

The first-level trigger required an energy deposit of at least 464~MeV 
in the electromagnetic section of RCAL (excluding the towers immediately 
around the beam pipe) and at least one track candidate in the CTD. Events 
with an energy deposit larger than 
1250~MeV in the FCAL towers surrounding 
the beam pipe were rejected in order to 
suppress proton beam-gas events along with a large fraction of other 
photoproduction events. No requirements were made on the LPS information.

At the second-level trigger, the background was reduced by using the 
measured times of the energy deposits and the summed energies from the 
calorimeter.

The full event information was available at the third-level trigger;
however, only a simplified reconstruction procedure was used.
Tighter timing cuts as well as algorithms to remove 
cosmic muons were applied. 
One  reconstructed vertex 
was demanded, with a $Z$ coordinate within $\pm 66$~cm of the nominal 
interaction point.  Furthermore, the events were required to satisfy at 
least one of the following conditions:
\begin{enumerate}
\item fewer than four reconstructed tracks and at least one pair with 
invariant mass less than 1.5~GeV (assuming they are pions); 
\item fewer than six reconstructed tracks and no pair with invariant mass 
larger than 5~GeV (again assuming pions).
\end{enumerate}

\noindent
Both sets of third-level triggers were prescaled by a factor six. 
Approximately $3 \times 10^5$ events were selected in this way, 
from an integrated luminosity of $898 \pm 14$~nb$^{-1}$ 
(the luminosity corresponding to no prescale).

\subsubsection{Offline requirements}
\label{offline}

After performing the full reconstruction of the events,
the following offline requirements were imposed to select elastic 
$\rho^0 \rightarrow \pi^+\pi^-$ candidates with a high momentum scattered
proton:

\begin{itemize}

\item Exactly two tracks in the CTD from particles of opposite charge, both 
associated with the reconstructed vertex.

\item The $Z$ coordinate of the vertex within $\pm30$~cm 
and the radial distance within 1.5~cm of the nominal interaction point.

\item In BCAL and RCAL, not more than 200 MeV in any EMC (HAC) 
calorimeter cell which is more than 30~cm (50~cm) away from the 
extrapolated impact position of either of the two tracks. This cut 
rejects events with additional particles, along with events with the
scattered positron in RCAL.


\item One track in the LPS with $0.98<x_L<1.02$. 
This corresponds to a $\pm 5\sigma$ window around $x_L=1$,
for an $x_L$ resolution of $0.4\%$.
As stated in section~\ref{alignment}, elastic photoproduction of $\rho^0$ mesons 
peaks at values 
of $x_L$ which differ from unity by less than $2 \times 10^{-3}$.
This requirement is used to tag elastic events.

\item Protons whose reconstructed trajectories come closer than 0.5~mm 
to the wall of the beam pipe, at any point between the vertex and the last 
station hit, were rejected. This eliminates events where the proton could have
hit the beam pipe wall and showered. In addition, it removes any
sensitivity of the acceptance to possible misalignments of the HERA beam
pipe elements.

\item The value of the $\chi^2/ndf$ of the fit to the proton track 
(cf. section~\ref{reconstruction}) less than 6.

\end{itemize}
\noindent 

The pion mass was assigned to each CTD track and the analysis was 
restricted to events reconstructed in the kinematic region defined by:
\begin{eqnarray}
                0.55 < & M_{\pi\pi}  & <  1.2 ~\mbox{GeV}, \nonumber \\
               0.27  < & p_T         & < 0.63 ~\mbox{GeV},  \nonumber \\
               50    < & W & < 100   ~\mbox{GeV},\\
                       & Q^2&< 1     ~\mbox{GeV}^2. \nonumber 
\label{kin}  
\end{eqnarray}
\noindent
The restricted range in the two-pion invariant mass $M_{\pi\pi}$ 
reduces the contamination from 
reactions involving other vector mesons, in particular from elastic 
$\phi$ and $\omega$ production. The limits on $p_T$, which is measured 
with the LPS, remove regions in 
which the acceptance of the LPS changes rapidly (cf. section~\ref{montecarlo}).

The photon-proton centre-of-mass energy $W$ and the mass $M_{\pi\pi}$ were determined from 
the momenta of the two pions~\cite{rho93}. 
Energy and momentum conservation relate the photon energy, 
$E_{\gamma}$, to the two-pion system energy $E_{\pi \pi}$ and 
longitudinal momentum $p_{Z\pi \pi}$ by 
$2E_{\gamma} \approx (E_{\pi \pi} - p_{Z\pi \pi})$, 
under the assumption that the positron emits the virtual photon with zero 
transverse momentum. Therefore $ W^2 \approx 4 E_\gamma E_p   \approx 
2 (E_{\pi \pi} - p_{Z\pi \pi}) E_p.$ 
From the Monte Carlo study discussed in the next section, the resolution on 
$W$ has been found to be about 2~GeV; that on $M_{\pi\pi}$ is about $30$~MeV.

\begin{figure}
\vspace{-1.0cm}
\begin{center}
\leavevmode
\hbox{%
\hspace*{-0.8cm}
\epsfxsize = 15cm
\epsffile{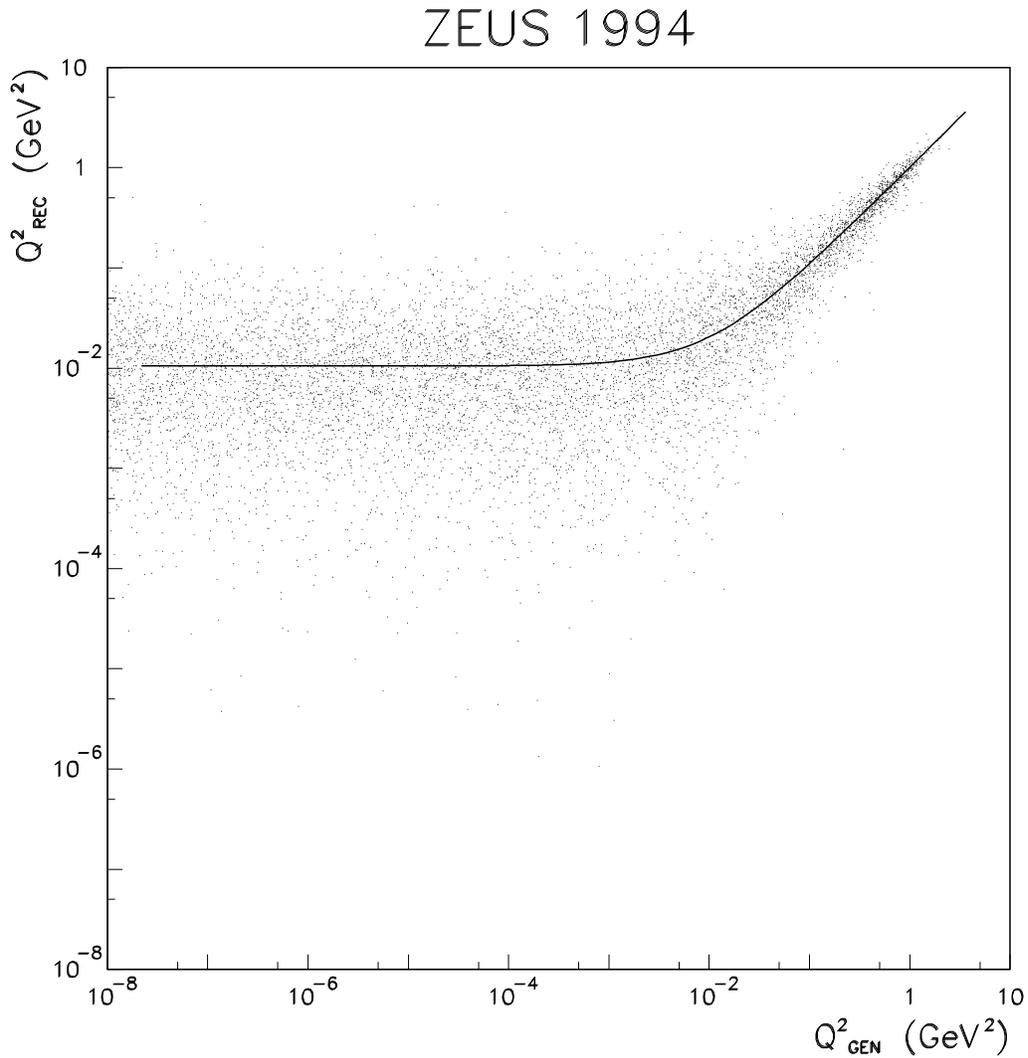}}
\end{center}
\vspace{-1cm}
\caption{Scatter plot of reconstructed versus generated values of $Q^2$
for Monte Carlo events. The continuous line shows the expected average relation 
assuming a spread of the beam transverse momentum with standard 
deviations $\sigma_{p_X}=40$~MeV and $\sigma_{p_Y}=90$~MeV in the horizontal and 
in the vertical directions, respectively. 
}
\label{figq2_resolution}
\end{figure}

The combination of the trigger requirements and the offline cuts, excluding
that on $Q^2$, limits $Q^2$ to be less than 
$\approx 4$~GeV$^2$. 
However, unlike previous ZEUS analyses of untagged photoproduction 
events, for the
present data $Q^2$ was determined event by event. By exploiting the 
transverse momentum balance of the scattered positron, the $\pi^+ \pi^-$ 
pair and the scattered proton, one obtains 
$p_{T}^e$, the transverse momentum of the scattered positron.
The variable $Q^2$ was then calculated from $Q^2=(p_{T}^e)^2/(1-y)$, where
$y$ is the fraction of the positron energy transferred
by the photon to the hadronic final state, in the proton rest frame; it 
was evaluated as $y \approx W^2/s$, where $\sqrt{s}$ 
is the $ep$ centre-of-mass energy.
Figure~\ref{figq2_resolution} shows a scatter plot of the reconstructed
and the generated value of $Q^2$ for the sample of Monte Carlo events 
used to evaluate the acceptance (cf. section~\ref{montecarlo}); the line 
shows 
the expected average relationship between these two quantities assuming 
a beam transverse momentum distribution with $\sigma_{p_X}=40$~MeV and 
$\sigma_{p_Y}=90$~MeV.
At small values of $Q^2$, the resolution on $Q^2$ is dominated by the beam 
transverse momentum spread; it is about 
100\% at $3 \times 10^{-2}$~GeV$^2$, 40\% at 0.1~GeV$^2$ and 
20\% at 1~GeV$^2$.

\begin{figure}
\vspace{-1.5cm}
\begin{center}
\leavevmode
\hbox{%
\hspace*{-0.8cm}
\epsfxsize = 15cm
\epsffile{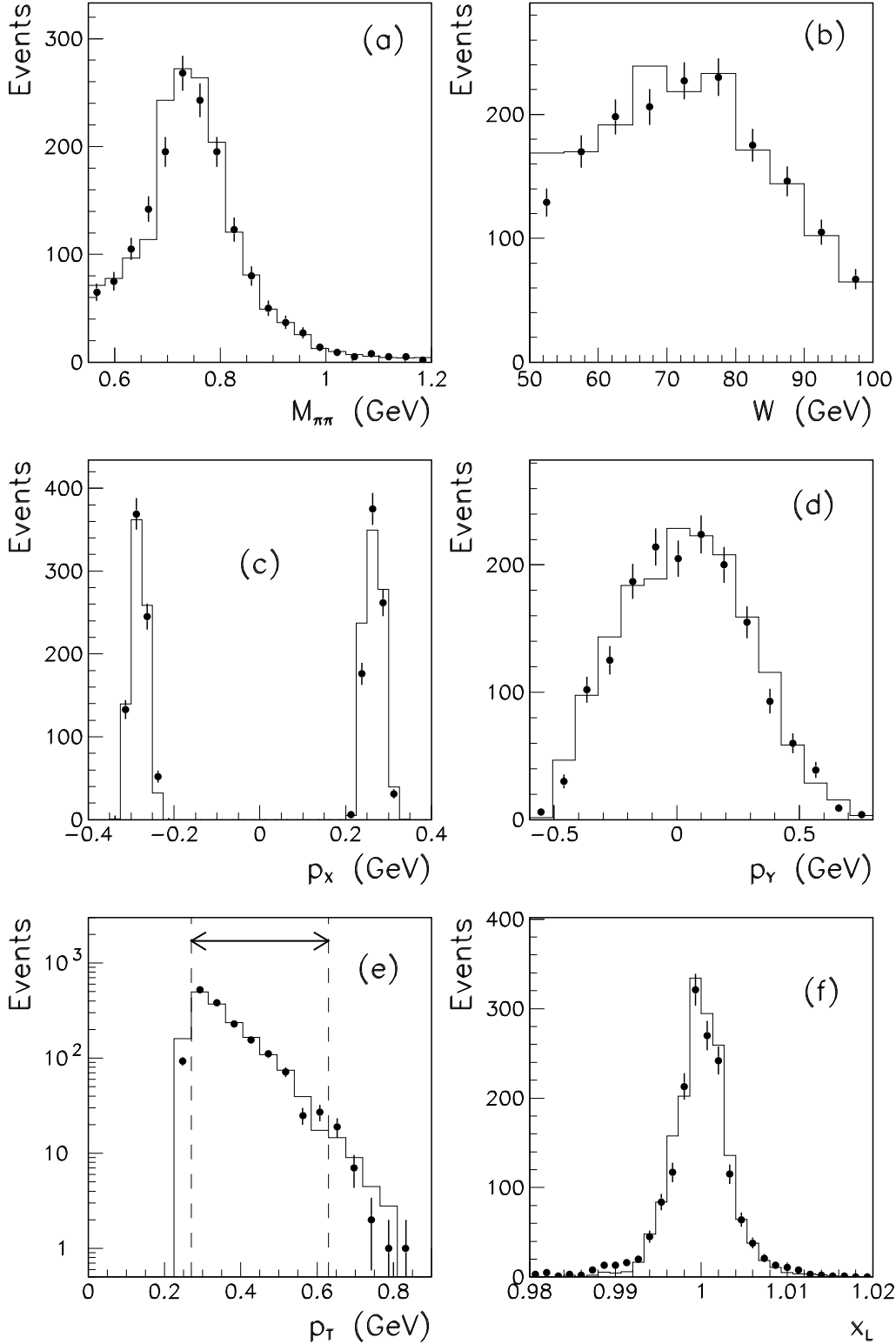}}
\end{center}
\vspace{-1cm}
\caption{Observed distributions for
(a) $M_{\pi\pi}$, (b) $W$, (c) $p_X$, (d) $p_Y$,
(e) $p_T$ and  (f) $x_L$  of
the reconstructed data (points) and the reconstructed Monte Carlo events 
(histogram). The distributions are not corrected for acceptance. The 
vertical bars indicate the statistical uncertainties. The dashed
lines in (e) show the limits of the $p_T$ region used.
}
\label{dataMC} 
\end{figure}

\begin{figure}
\vspace{-1.5cm}
\begin{center}
\leavevmode
\hbox{%
\hspace*{-0.8cm}
\epsfxsize = 15cm
\epsffile{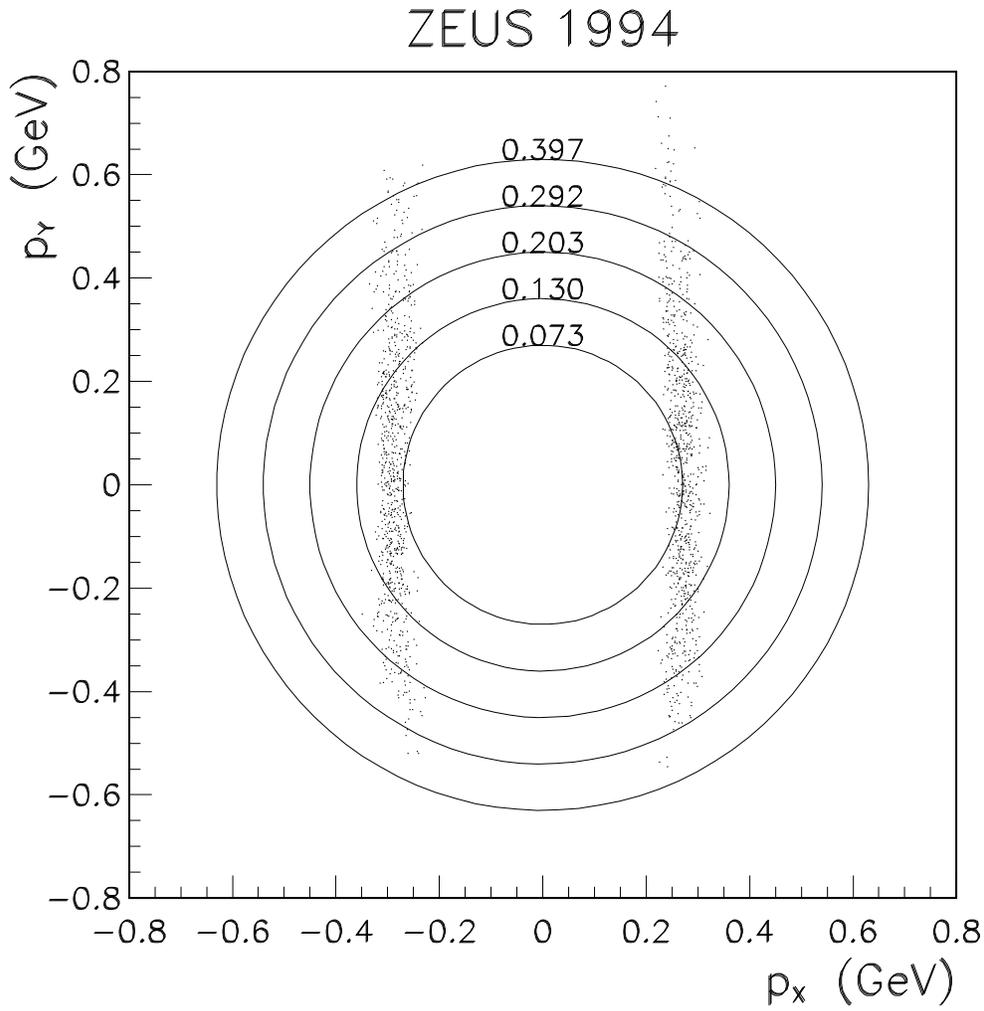}}
\end{center}
\vspace{-1cm}
\caption{Scatter plot of the $X$ and $Y$ components of the scattered 
proton momentum as measured by the LPS for the accepted events. The 
continuous curves correspond to the indicated values of $|t|$ 
(in GeV$^2$), which were used as limits of the bins in 
Fig.~\protect\ref{dndt}.
}
\label{pxvspy}
\end{figure}

The final sample contains 1653 events. 
Figure~\ref{dataMC} shows the 
$M_{\pi\pi}$, $W$, $p_X$, $p_Y$, $p_T$ and $x_L$ 
distributions after the offline selections; the variables $p_X$ and $p_Y$ 
denote the transverse components of the outgoing proton momentum with respect to the
incoming beam axis and $p_T^2=p_X^2+p_Y^2$. The invariant mass plot
is dominated by the $\rho^0$ peak. The shape of the $p_X$ spectrum, with 
two well separated peaks, is a 
consequence of the fact, discussed earlier, that events with $x_L$ close 
to unity only populate a narrow region of the detectors at $v\approx 0$.
As discussed earlier, elastic $\rho^0$ photoproduction  peaks at $x_L=1$ to
within $< 2 \times 10^{-3}$; the width of the $x_L$ distribution in 
Fig.~\ref{dataMC}f shows that the resolution of the LPS in $x_L$ 
is $\approx 0.4\%$. For the same events,
Fig.~\ref{pxvspy} shows the scatter plot of the reconstructed $X$ and 
$Y$ components of the proton momentum.
For the present measurement,
only the $p_T$ region between the dashed
vertical lines in Fig.~\ref{dataMC}e was used.
The variable $t=(p-p^{\prime})^2$, where $p$ and 
$p^{\prime}$ are the incoming and the scattered proton four-momenta,
respectively, can be evaluated as follows: 
\begin{eqnarray}
t=(p-p^{\prime})^2 \approx -\frac{p_T^2}{x_L} 
\left[1+(M_p^2/p_T^2)(x_L-1)^2\right],
\end{eqnarray}
\noindent
where $M_p$ is the proton mass and terms of order
$(M_p^2/\vec{p^{\prime}}^2)^2$ or higher are neglected.
For the present events, which have $x_L\approx 1$, the approximation
$t\approx -p_T^2/x_L\approx -p_T^2$ was used. 

\begin{figure}
\vspace{-1.0cm}
\begin{center}
\leavevmode
\hbox{%
\hspace*{-0.8cm}
\epsfxsize = 15cm
\epsffile{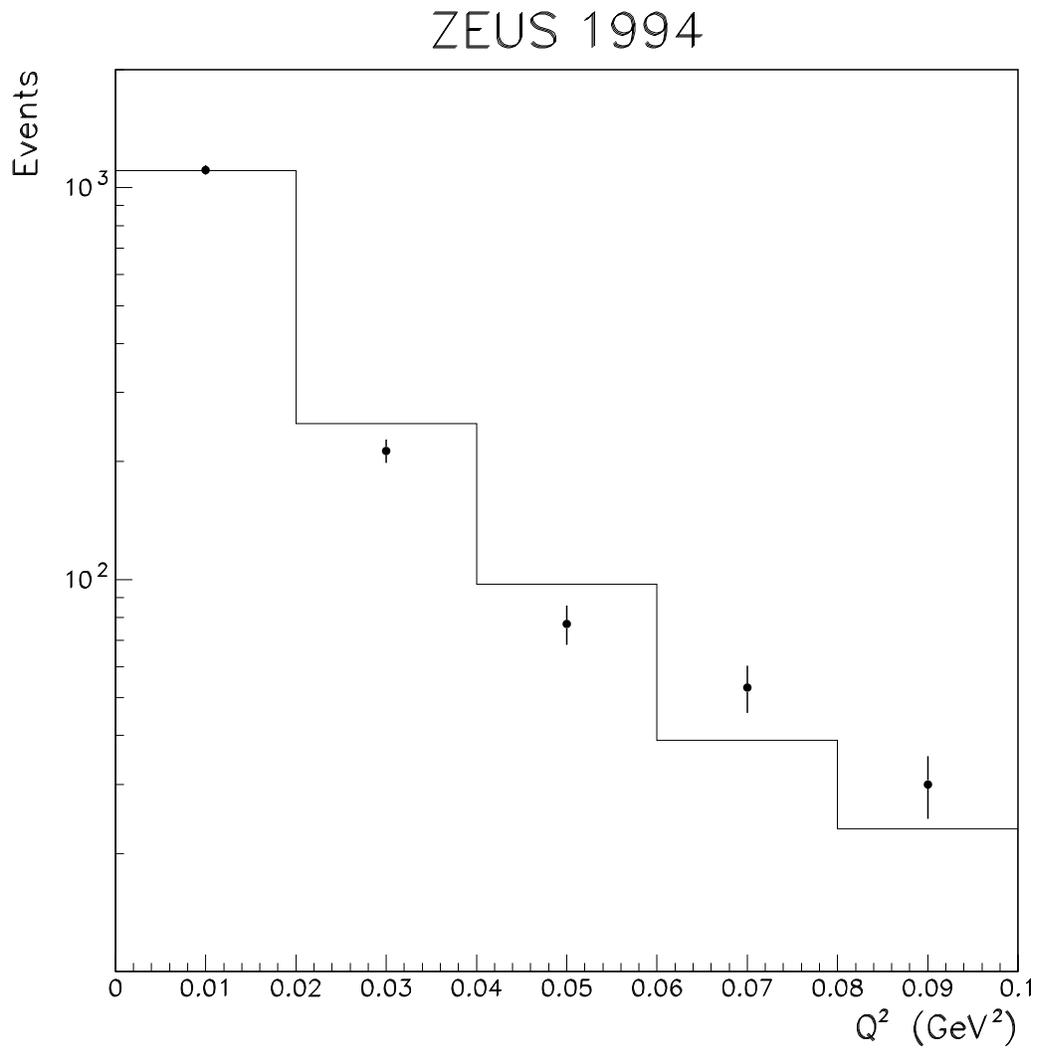}}
\end{center}
\vspace{-1cm}
\caption{$Q^2$ distributions of
the reconstructed data (points) and the reconstructed Monte Carlo events 
(histogram). The distributions are not corrected for acceptance. 
Only the region $Q^2<0.1$ is shown. The 
vertical bars indicate the size of the statistical uncertainties. 
}
\label{q2}
\end{figure}

Finally, Fig.~\ref{q2} shows the distribution of the reconstructed
values of $Q^2$. As discussed above, at small values of $Q^2$ the 
intrinsic spread of the beam transverse momentum dominates. The requirement
that $Q^2$ be less than 1 GeV$^2$ removes 7 events.
The median $Q^2$ of the data, estimated with the Monte Carlo study
discussed in the next section, is approximately $10^{-4}$~GeV$^2$.

\subsection{Monte Carlo generators and acceptance calculation}
\label{montecarlo}

The reaction $ep \rightarrow e\rho^0 p$ was modelled using the 
DIPSI~\cite{dipsi} generator, which was shown to reproduce the ZEUS 
$\rho^0$ photoproduction data~\cite{rho93}. The effective $W$ 
dependence of the $\gamma p$ cross section for the events generated
was of the type $\sigma \propto W^{0.2}$. The $t$ distribution was 
approximately exponential with a slope parameter of 9.5~GeV$^{-2}$. 
The two-pion invariant mass, $M_{\pi\pi}$, was generated so as to 
reproduce, after reconstruction,  the measured distribution.  The 
angular distribution of the decay pions was assumed to be that expected 
on the basis of $s$-channel helicity conservation~\cite{shilling-wolf}.

\begin{figure}
\vspace{-1.5cm}
\begin{center}
\leavevmode
\hbox{%
\hspace*{-0.8cm}
\epsfxsize = 15cm
\epsffile{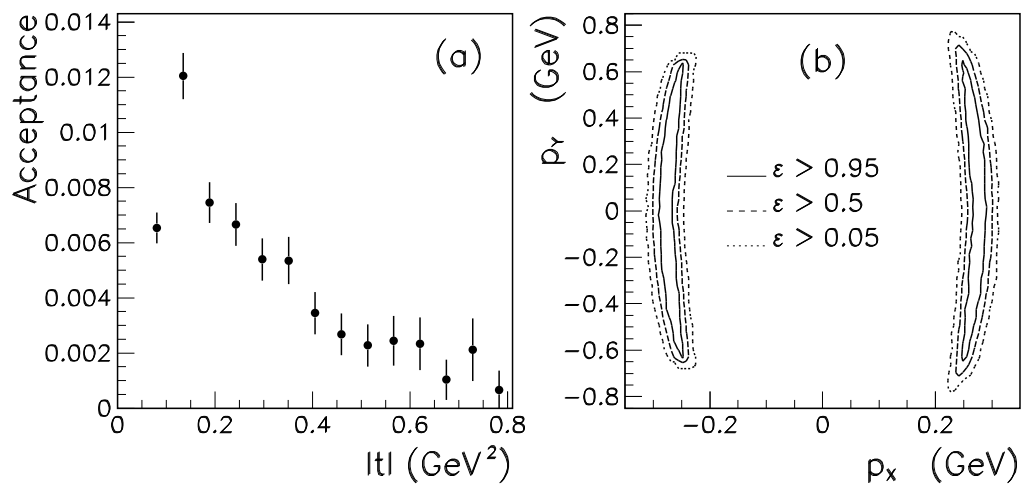}}
\end{center}
\vspace{-8cm}
\caption{(a) Acceptance as a function of $t$; it
includes the effects of 
the geometric acceptance of the apparatus (ZEUS and LPS), of its efficiency 
and resolution and of the trigger and reconstruction efficiencies. 
The vertical bars indicate the size of the statistical uncertainties.
(b) Purely geometric acceptance $\varepsilon$ for the LPS alone 
for $x_L=1$ tracks as a function of $p_X$ and $p_Y$. 
}
\label{acceptance}
\end{figure}

The simulated events were passed through the same reconstruction and 
analysis programs as the data. In Figures~\ref{dataMC} and~\ref{q2}
the distributions of the reconstructed data (not corrected for acceptance) 
over $M_{\pi \pi}$, $W$, $p_X$, $p_Y$, $p_T$, $x_L$ and $Q^2$ 
are compared with those obtained for the reconstructed Monte Carlo events.
The Monte Carlo is in reasonable agreement with the data.
Figure~\ref{acceptance}a shows the overall acceptance as a function 
of $t$, obtained using DIPSI. The acceptance includes the effects of 
the geometric acceptance of the apparatus, its efficiency and 
resolution and the trigger and reconstruction efficiencies. 
Since the detector planes cannot be positioned in the beam,
the acceptance vanishes at small values of $t$.
Conversely, in the $p_X, p_Y$ region covered by the detectors, 
the acceptance of the LPS is large, as shown in Fig.~\ref{acceptance}b,
which shows the geometric acceptance of the LPS alone, irrespective
of the acceptance of the rest of the ZEUS apparatus. 
The region of LPS geometric acceptance larger 
than 95\% for both $p_X>0$ and $p_X<0$ maps into that of 
$0.25 \lsim p_T \lsim 0.65$~GeV;
as discussed in section~\ref{offline}, events outside this region
are not used in the present analysis. 
For events with elastic $\rho^0$ photoproduction, the geometric 
acceptance of the LPS, averaged over azimuth, is approximately 
6\%. 

As discussed in the next section, in order to estimate the 
contamination from the reaction 
$ep \rightarrow e  \rho^0 X_N$, where $X_N$ is 
a hadronic state of mass $M_N$ resulting from the diffractive 
dissociation 
of the proton, the PYTHIA generator~\cite{pythia} was used. 
A cross section of the form $d^2\sigma /dt dM_N^2 \propto e^{-b|t|}/M_N^2$, with 
$b=6$ GeV$^{-2}$, was assumed; the value of $M_N$ ranged between 
$M_p+M_{\pi}$ (where $M_p$ is the proton and $M_{\pi}$ the pion mass)
and a maximum fixed by the condition  $M_N^2/W^2 \le 0.1$~\cite{chapin}. 
The $\rho^0$ decay 
angular distributions were assumed to be 
the same as those of the elastic events.

\subsection{Backgrounds}
\label{backgrounds}

\begin{figure}
\vspace{-1.5cm}
\begin{center}
\leavevmode
\hbox{%
\hspace*{-0.8cm}
\epsfxsize = 15cm
\epsffile{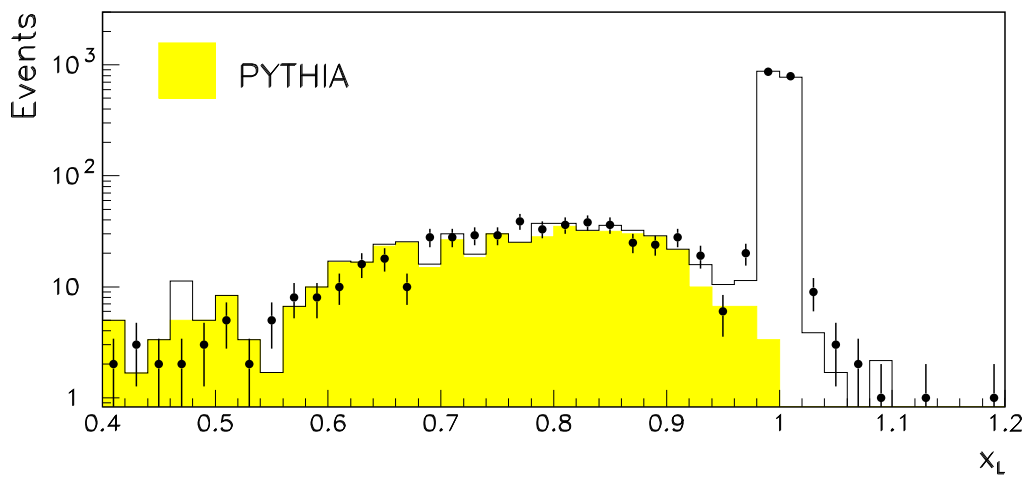}}
\end{center}
\vspace{-8cm}
\caption{Reconstructed $x_L$ spectrum for the data (points).
All selection cuts were applied except that on $x_L$.
The distribution is not corrected for acceptance.
The sum of the reconstructed $x_L$ distributions produced with
the DIPSI and PYTHIA generators was fitted to this spectrum with 
the normalisations as 
free parameters of the fit. The result of the fit is shown as a histogram. 
The DIPSI generator simulates the elastic reaction $ep \rightarrow e \rho^0 p$,
and PYTHIA the proton dissociative reaction $ep \rightarrow e \rho^0 X_N$.
The contribution of PYTHIA is shown as the hatched area. 
The vertical bars indicate the size of the statistical uncertainties.
} 
\label{pdiffr} 
\end{figure}

After applying the selection criteria described in 
section~\ref{event_selection},
the data contain small contributions from various background 
processes to the reaction $ep \rightarrow e \pi^+ \pi^- p$:

\begin{itemize}

\item Beam-halo tracks observed in the LPS may overlap with events in which 
a $\rho^0$ is seen by 
the ZEUS central apparatus. 
The term beam-halo refers to protons with 
energy close to that of the beam originating from 
interactions of beam protons with the residual gas in the
pipe or with the beam collimators.
Such tracks are completely uncorrelated with the activity
in the central ZEUS apparatus; therefore, 
any sample of events selected without using the LPS information 
contains the same fraction, $\epsilon_{halo}$, of random overlaps from 
halo tracks within 
the LPS acceptance. This fraction was found to be 
$\epsilon_{halo}=0.25 \pm 0.03\%$ by analysing
events of the type $ep \rightarrow eX$ at $Q^2 > 4$~GeV$^2$ in 
which the virtual photon diffractively dissociates into the state $X$.
For these events one can measure $X$ and the scattered positron
in the calorimeter; in addition a proton track is looked for in the LPS.
If one is found, the event is fully contained and its
kinematics is thus overconstrained: most beam-halo events 
appear to violate energy-momentum conservation and can therefore 
be identified.

The contamination of 
the present sample (after the requirement of a good LPS track) 
can be obtained as 
$(\epsilon_{halo} N_{no LPS})/N_{LPS}=5.0\% \pm 0.6\%~(\mbox{stat.})$, where 
$N_{no LPS}$ indicates the number of events 
found by applying all cuts except for the requirement that a track be
detected in the LPS, and $N_{LPS}=1653$ is the number of events 
after all cuts. 
These events were not removed from the present
sample, but their effect on the measurement of the $t$ slope is small, as 
discussed in section~\ref{results}.

\item
In the reaction $ep \rightarrow e  \rho^0 X_N$,
the proton diffractively dissociates into a hadronic 
state $X_N$ which may escape detection by the central detector. 
The debris 
of $X_N$ may contain a proton within the LPS acceptance and with
$0.98<x_L<1.02$: such events
are indistinguishable from elastic $\rho^0$ production.

In order to evaluate the contamination from such events, the cut on 
$x_L$ was removed; Fig.~\ref{pdiffr} shows the $x_L$ spectrum thus 
obtained, not corrected for acceptance. 
The sum of the reconstructed $x_L$ distributions 
from DIPSI and PYTHIA was fitted to this spectrum with 
the normalisations of the simulated distributions as 
free parameters of the fit.
The fit gives an acceptable description of the data, as shown in 
Fig.~\ref{pdiffr}. The resulting 
contamination of proton-dissociative events for $x_L>0.98$ is 
$0.21\% \pm 0.15\%~(\mbox{stat.})$, a major improvement with 
respect to $11\% \pm 1\%~(\mbox{stat.}) 
\pm 6\%~(\mbox{syst.})$ in the earlier ZEUS result~\cite{rho93} which did not
use the LPS.

\item
The contaminations from elastic production of $\omega$ and $\phi$ mesons 
were estimated in~\cite{rho93} to be $(1.3\pm0.2)\%$ and
$(0.3 \pm 0.1)\%$, respectively. 

\item

Contamination from positron beam-gas and proton beam-gas events was 
studied by using the unpaired bunches event samples to which all the cuts 
described above were applied. No event passed the cuts, indicating a 
negligible contamination.

\end{itemize}

\section{Results}
\label{results}

The differential cross section $d\sigma/dM_{\pi\pi}$ for the process
$\gamma p \rightarrow \pi^+ \pi^- p$ was evaluated 
in the kinematic range
$0.55<M_{\pi\pi}<1.2$~GeV, $50<W<100$~GeV, $Q^2<1$~GeV$^2$ and 
$0.073<|t|<0.40$~GeV$^2$.
In each bin the 
cross section was obtained as

\begin{eqnarray}
\frac {N_{\pi^+\pi^-}}{a L \Phi}c_{halo},
\label{crosssection}
\end{eqnarray}

\noindent
where $N_{\pi^+\pi^-}$ is the number of observed events in the bin,
$L$ is the integrated luminosity, 
$a$ is the overall acceptance in the bin, 
and $\Phi=0.0574$ is the photon flux factor, i.e. the integral of
equation (5) in ref.~\cite{rho93} over the measured $W$ and $Q^2$ ranges
of this measurement. The factor $c_{halo}=0.950 \pm 0.006$~(stat.) corrects for 
the beam-halo contamination discussed in section~\ref{backgrounds}. 

The effects of positron initial and final state radiation 
and that of vacuum polarisation loops were neglected; the effects
on the integrated cross section were estimated to be smaller 
than 4\%~\cite{kurek}.
The effects on the shape of the $M_{\pi\pi}$ and $t$ distributions are 
expected to be negligible. The small residual contaminations 
from proton dissociative
$\rho^0$ production and elastic $\omega$ and $\phi$ production,
discussed in the previous section, were not corrected for.

\begin{figure}
\vspace{-1.5cm}
\begin{center}
\leavevmode
\hbox{%
\hspace*{-0.8cm}
\epsfxsize = 15cm
\epsffile{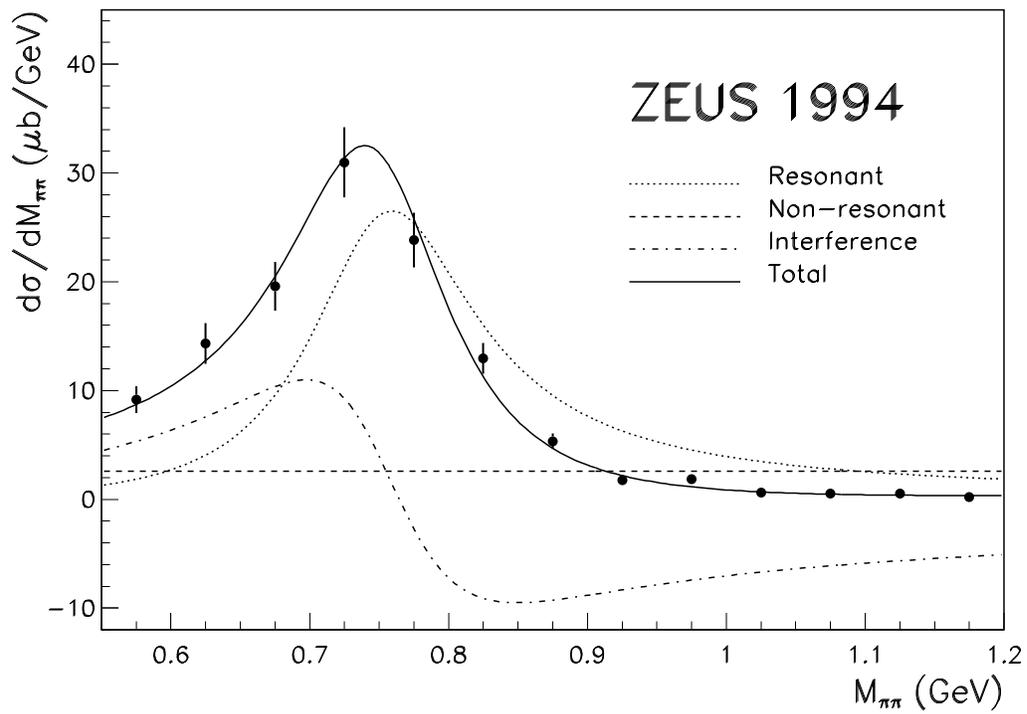}}
\end{center}
\caption{Differential cross section $d\sigma/dM_{\pi\pi}$ 
for the reaction $\gamma p \rightarrow \pi^+\pi^- p$,
in the region $0.55<M_{\pi\pi}<1.2$~GeV and for
$\langle W \rangle = 73$~GeV. 
The vertical bars indicate the size of the statistical uncertainties only.
The lines indicate the result of the fit using expression 
(11) of ref.~\protect\cite{rho93}.
} 
\label{rec_mass} 
\end{figure}

Figure~\ref{rec_mass} shows the differential cross section 
$d\sigma/dM_{\pi\pi}$ in the interval
$0.55< M_{\pi\pi} < 1.2$~GeV, $0.073<|t|<0.40$~GeV$^2$ for
$\langle W \rangle = 73$~GeV. The mass spectrum is skewed, as previously 
observed
both at low energy and at HERA. This can be understood 
in terms of resonant and
non-resonant $\pi^+\pi^-$ production~\cite{drell}, and their 
interference~\cite{soeding}. The spectrum was fitted using expression (11)
of ref.~\cite{rho93}. The results for the total, resonant and interference 
terms, as obtained in the fit, are indicated in the figure.
The fraction of the resonant to the total contribution in the measured range
was found to be $c_{res}=0.91 \pm 0.04$~(syst.).
The uncertainty was evaluated by repeating the fit with 
the various functional forms discussed in~\cite{rho93}.
In~\cite{rho93} the contribution of the resonant term was found to 
vary from $86\%$ for $|t|=0.01$~GeV$^2$ to $95\%$ for $|t|=0.5$ GeV$^2$.
No $t$ dependence of $c_{res}$ was assumed here, except in the 
evaluation of the systematic uncertainty (see below).

\begin{figure}
\vspace{-1.0cm}
\begin{center}
\leavevmode
\hbox{%
\hspace*{-0.8cm}
\epsfxsize = 15cm
\epsffile{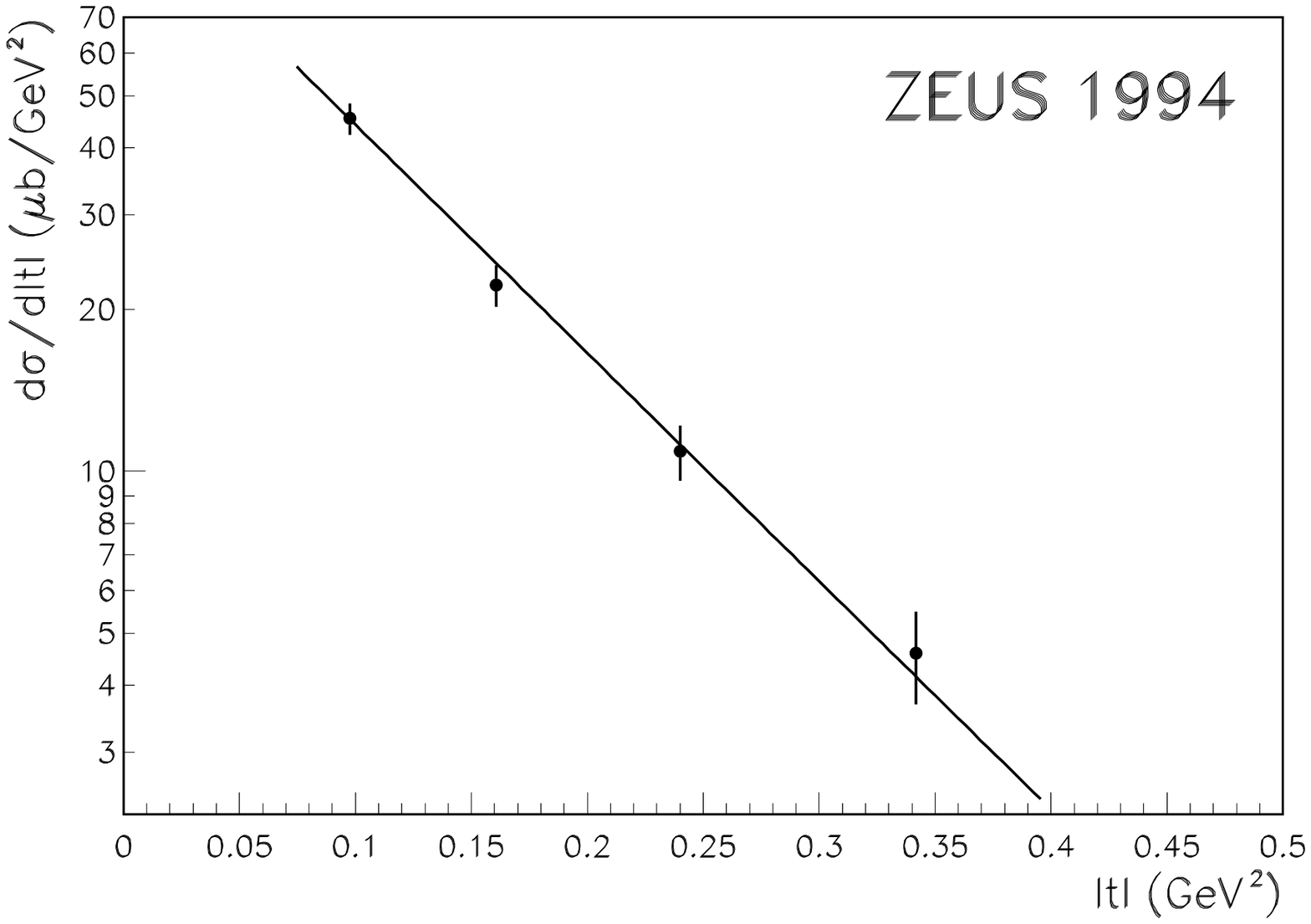}}
\end{center}
\vspace{-1cm}
\caption{Differential cross section $d\sigma/d|t|$ 
for elastic $\rho^0$ photoproduction, $\gamma p \rightarrow \rho^0 p$,
at $\langle W \rangle=73$~GeV in the region $0.073<|t|<0.40$~GeV$^2$,
$0.55<M_{\pi\pi}<1.2$~GeV.
The vertical bars indicate the size of the statistical uncertainties only.
The line is the result of the fit described in the text. 
}
\label{dndt}
\end{figure}

The differential cross section $d\sigma/dt$ for the reaction
$\gamma p \rightarrow \rho^0 p$ was obtained similarly to
$d\sigma/dM_{\pi\pi}$, but in addition the correction factor $c_{res}$, just
discussed, was applied.
Figure~\ref{dndt} shows the result 
in the interval $0.073<|t|<0.40$~GeV$^2$, $0.55< M_{\pi\pi} < 1.2$~GeV 
for $\langle W \rangle = 73$~GeV.
The data were fitted with the function 

\begin{eqnarray}
\frac{ d\sigma}{dt}   = A     \cdot e^{-b |t|};
\label{single}
\end{eqnarray}

\noindent
the result of the fit is shown as a straight line on Fig.~\ref{dndt}.
The fitted value of the slope parameter $b$ is 

\begin{eqnarray}
b = 9.8\pm 0.8~(\mbox{stat.})
\pm 1.1~(\mbox{syst.})~\mbox{GeV}^{-2}.
\label{result}
\end{eqnarray}

\noindent
The result is consistent with 
$b=9.9 \pm 1.2~(\mbox{stat.}) \pm 1.4~(\mbox{syst.})$~GeV$^{-2}$ 
obtained in~\cite{rho93}
for the range $60<W<80$~GeV, $Q^2<4$~GeV$^2$ and $|t|<0.5$~GeV$^2$ using
a fit of the type $A\exp{(-b|t|+ct^2)}$. For both the present data and 
those of ref.~\cite{rho93}, $\langle W \rangle \approx 70$~GeV.

The measured differential cross section was integrated over the range
$0.073<|t|<0.40$~GeV$^{-2}$, yielding 
$\sigma = 5.8 \pm 0.3~(\mbox{stat.})\pm 0.7~(\mbox{syst.})~\mu\mbox{b}$, 
again at $\langle W \rangle =73$~GeV and for
$0.55< M_{\pi\pi} < 1.2$~GeV. The result can be extrapolated to the mass range
$2M_{\pi}< M_{\pi\pi} < M_{\rho}+5 \Gamma_0$, as in~\cite{rho93},
using the fit to the mass spectrum described earlier (here $\Gamma_0$ is the
$\rho^0$ width); this yields   
$\sigma = 6.3 \pm 0.3~(\mbox{stat.})\pm 0.8~(\mbox{syst.})~\mu\mbox{b}$,
where no uncertainty was assigned to the extrapolation. If our previous
result~\cite{rho93} is integrated in the $t$ range covered by the 
present data, using the published results of the fit with the function
$A\exp{(-b|t|+ct^2)}$ (table 5 of ref.~\cite{rho93}, left column),
 one finds 
$\sigma = 6.7 \pm 1.1~(\mbox{syst.})~\mu\mbox{b}$, in good agreement
with the present result; only the systematic uncertainty is given since it is 
dominant. 
This uncertainty was
obtained by scaling the one published in~\cite{rho93} for the 
cross section measured over the range $|t|<0.5$~GeV$^2$ by the ratio 
of the cross sections for the present $t$ range and for $|t|<0.5$~GeV$^2$.

The major sources of systematic uncertainty on $b$ and $\sigma$ 
are the acceptance 
determination and the background contamination, the former being 
dominant. Table~\ref{systematics} lists the individual contributions.
In the following we discuss them in detail.

\begin{table}
\begin{center}
\begin{tabular}{lcc}                          \hline \hline
 Contribution                             &  $|\Delta b/b|$ & $|\Delta \sigma/\sigma|$ \\ \hline 
 Integrated luminosity                    &    -            & 1.5\%  \\ 
 Acceptance: trigger efficiency           &    -            & 9\%  \\
 Acceptance for pion tracks               &    $<1\%$       & 1\%  \\
 Acceptance for proton track              &    7\%          & 6\%     \\
 Acceptance: sensitivity to binning in $t$&    2\%          &   -  \\
 Acceptance: unfolding of beam transverse
             momentum spread              &     7\%         &   -  \\
 Acceptance: sensitivity of $p$ beam angle&     3\%         &  1\% \\
 Background: beam-halo                    &     4\%         &   -  \\           
 Procedure to extract the resonant 
 part of the cross section                &     1.6\%       & 4\% \\
 Background due to elastic 
 $\omega$ and $\phi$ production           &    -            &  1\%  \\
 Radiative corrections                    & -               & 4\%   \\
 \hline
 Total                                    & 11\%            & 12\%    \\ 
 \hline \hline
\end{tabular}
\end{center}
\caption{Contributions to the systematic uncertainty on $b$ and $\sigma$.
}
\label{systematics}
\end{table}

\begin{enumerate}

\item 
In order to estimate the uncertainty due to the acceptance, 
the analysis was repeated varying the requirements and 
procedures as listed below.

\begin{enumerate} 

\item For the pion tracks in the central detector:
\begin{itemize}
\item The pseudorapidity $\eta=-\ln{\tan{(\theta/2)}}$
of each of the two tracks was restricted
to the range $|\eta|~<~1.8$, thereby using only tracks which 
have traversed at least three superlayers in the CTD.


\item The radial distance of the vertex from the beam axis was required 
to be less than 1~cm.

\end{itemize}

In both cases the changes are small; by summing them in quadrature
one finds $|\Delta b/b|=0.2\%$ and $|\Delta \sigma/\sigma|=1\%$. 

\item For the proton track in the LPS:

\begin{itemize}

\item The maximum allowed value of $\chi^2/ndf$ for the reconstructed proton 
track was reduced from 6 to 2.

\item The minimum distance of approach of the proton 
trajectory to the beam pipe was increased from 0.5~mm to 
1.5~mm.  

\item Events with $p_X>0$ and with $p_X<0$ were analysed separately, 
as a check of possible relative rotations of the stations.

\item The data were divided into a ``large acceptance" and a ``low 
acceptance" sample depending on the position of the LPS stations, 
which, as discussed above, varied slightly from run to run. 

\end{itemize}

By summing the individual contributions to $\Delta b/b$ in quadrature, 
independently of their sign, $|\Delta b/b|=7\%$ is obtained. The corresponding
uncertainty on $\sigma$ is $|\Delta \sigma/\sigma|=6\%$. 

\item The sensitivity of the result on $b$ to the binning in $t$ was studied by 
reducing bin sizes by up to 20\%; the bin edges were moved by up to one 
fourth of the bin size. The largest effect was $2\%$ for
$|\Delta b/b|$.

\item As discussed earlier, $t$ has been obtained as $-p_T^2$, with
$p_T^2=p_X^2+p_Y^2$ the transverse momentum of the scattered proton
with respect to the incoming beam axis. Since the incoming proton beam has an
intrinsic transverse momentum spread of $\sigma_{p_X} \approx 40$~MeV and
$\sigma_{p_Y} \approx 90$~MeV, which is much larger than the LPS resolution
in transverse momentum, the measured value of $t$ is smeared with
respect to the true $t$.
The Monte Carlo simulation takes into account the 
proton beam transverse momentum spread. 
The acceptance corrected $t$ distribution is thus corrected 
also for this effect. 

The following alternative approach to account for the effect of
the transverse momentum spread of the beam has also been followed.
Assuming that the true $t$ distribution has the form given 
by equation~(\ref{single}), 
the measured $p_T^2$ distribution can be expressed as a
convolution of equation~(\ref{single}) and a two-dimensional Gaussian 
distribution 
representing the beam transverse momentum distribution, with standard
deviations $\sigma_{p_X}$ and $\sigma_{p_Y}$.
Unfolding the contribution of the beam transverse 
momentum spread from $d\sigma/dp_T^2$ provides an alternative evaluation 
of $d\sigma/dt$. In this case one first measures the distribution of
$p_T^2$ without making any correction for
the effects of the beam intrinsic spread, thereby 
exploiting the good resolution of the LPS on the transverse momentum.
In a second stage, the effect of the beam spread is unfolded.
If $\sigma_{p_X}=40$~MeV and $\sigma_{p_Y}=90$~MeV (as seen in the data, 
cf. Fig.~\ref{ctdlps}), 
then $|\Delta b/b|=7\%$, with 
only a weak dependence on the values of $\sigma_{p_X}$ and $\sigma_{p_Y}$.

\item The sensitivity to the determination of the proton beam angle (cf. 
section~\ref{alignment}) was evaluated by 
systematically shifting $p_T$ by 10~MeV. This amount is
twice the $p_T$ resolution of the LPS and corresponds 
to $>5$ times the uncertainty on the means of the distributions
of Fig.~\ref{ctdlps}. The corresponding variations
of $b$ and $\sigma$ were $|\Delta b/b|=3\%$ and 
$|\Delta \sigma/\sigma|=1\%$. 

\end{enumerate}

\noindent
The differences between the values of $b$ obtained in cases (a) to (e)
and that obtained with 
the standard analysis were summed in quadrature, yielding 
$|\Delta b/b|=10.5\%$ and $|\Delta \sigma/\sigma|=6\%$. 

\item
Effect of background contamination.

\begin{enumerate}
\item As mentioned above, no correction was applied for a possible $t$ 
dependence of the background. The only significant background 
is the halo. If the assumption is made that the halo contribution 
($5.0\% \pm 0.6\%$) has 
a distribution of the type $\exp{(-b_{halo}|t|)}$, then 
$|\Delta b/b|<4\%$ when $b_{halo}$ is varied
between 5 and 15~GeV$^{-2}$; this range of variation is 
consistent with estimates of $b_{halo}$ based on 
the $ep \rightarrow eXp$ events at $Q^2>4$~GeV$^2$ discussed in 
section~\ref{backgrounds}.

\item 
If the $t$ dependence of $c_{res}$ evaluated in~\cite{rho93} 
is assumed for the present data, the slope 
changes by $\Delta b/b=-1.6\%$.  

\end{enumerate}
\end{enumerate}

\noindent
The latter two contributions were also added quadratically to the 
systematic uncertainty, yielding a total systematic uncertainty
of 11\% on $b$, dominated by the LPS acceptance and the effect of the
beam transverse momentum spread.

The total systematic uncertainty on $\sigma$ is 12\%, which includes, 
in addition to the contributions detailed above,
the uncertainty on the luminosity (1.5\%), that 
on the trigger efficiency~\cite{rho93} (9\%) and that related to 
the extraction of the resonant part of the cross section.
The estimated background due to elastic $\omega$ and $\phi$ production,
as well as the upper limit of the correction for radiative effects have 
also been included. The systematic uncertainty on $\sigma$ is dominated 
by contributions not related to the LPS (11\%); the uncertainty
on the LPS acceptance is 6\%, which has only a small effect on the total 
uncertainty when summed in quadrature with the other contributions.

\section{Conclusions}

The Leading Proton Spectrometer of ZEUS is a large scale system of 
silicon micro-strip detectors which have been successfully operated
close to the HERA proton beam (typically a few mm) by 
means of the ``Roman pot" 
technique. It measures precisely the momentum of high energy scattered protons,
with accuracies of 0.4\% for the longitudinal and 5~MeV for the 
transverse momentum.

As a first application, the cross section, the $M_{\pi\pi}$ and the
$t$ dependences of the reaction 
$\gamma p \rightarrow \rho^0 p$ 
have been measured in the kinematic range $Q^2 <1$~GeV$^2$, 
$50<W<100$~GeV, $0.55< M_{\pi\pi}<1.2$~GeV and $0.073<|t|<0.40$~GeV$^2$.
Elastic events were tagged by demanding that $x_L$ be larger than 0.98,
i.e. that the scattered proton carry at least 98\% of the incoming 
proton momentum.
For the first time at these energies, $t$ was measured directly.
Compared to our previous analysis, the present technique 
based on the use of the LPS eliminates 
the contamination from events with diffractive dissociation 
of the proton into low mass states. 

In the range $0.073<|t|<0.40$~GeV$^2$, the differential cross section
$d\sigma/dt$ is described by an exponential distribution with a slope 
parameter 
$b = 9.8\pm 0.8~(\mbox{stat.}) \pm 1.1~(\mbox{syst.})~\mbox{GeV}^{-2}$.
The systematic uncertainty is dominated by the uncertainty on the 
LPS acceptance and the 
effect of the intrinsic transverse momentum spread of the beam.
In the measured $t$ and $M_{\pi\pi}$ intervals, the integrated 
$\rho^0$ photoproduction cross section at 
$\langle W \rangle =73$~GeV was found to be
$5.8\pm 0.3~(\mbox{stat.}) \pm 0.7~(\mbox{syst.})~\mu$b, consistent with our
previous measurement~\cite{rho93} obtained in a slightly different kinematic
range.


\section{Acknowledgements}

We thank the DESY directorate for their strong support and encouragement. 
We are also very grateful to the HERA machine group: collaboration 
with them was crucial to the successful installation and operation 
of the LPS.

We also want to  express our gratitude to all those who have participated
in the construction of the LPS, in particular to the very 
many people from the University of Bologna and INFN Bologna~($B$), CERN~($C$),
the University of Calabria and INFN Cosenza~($Cs$), DESY~($D$), 
LAA~($L$)~\cite{laa}, the University of Torino and INFN Torino~($T$), the
University of California at Santa Cruz~($S$),
who have so greatly contributed to the LPS project at various stages.

\noindent
For financial support we are grateful to the Italian
Istituto Nazionale di Fisica Nucleare (INFN), to the LAA project and
to the US Department of Energy. 

\noindent
For the mechanical design and construction of the stations and their
interface with HERA:  G.~Alfarone$^T$, F.~Call\`a$^T$, 
J.~Dicke$^D$, G. Dughera$^T$, P. Ford$^L$, 
H. Giesenberg$^D$, R.~Giesenberg$^D$, G. Giraudo$^T$, 
M.~Hourican$^L$, A. Pieretti$^B$, P. Pietsch$^D$. 

\noindent
For discussions on the optical design and on the mechanical interface with HERA
and for making some modifications to the machine layout to improve the LPS
acceptance:
W. Bialowons$^D$, R.~Brinkman$^D$, 
D.~Degele$^D$, R.~Heller$^D$, B.~Holzer$^D$,
R. Kose$^D$, M. Leenen$^D$, 
G. Nawrath$^D$, K.~Sinram$^D$, D. Trines$^D$, T.~Weiland$^D$ and F. Willeke$^D$.

\noindent
For advice on radiation doses and hardness of service electronics:
H. Dinter$^D$, B.~Lisowski$^B$ and H. Schoenbacher$^C$.  

\noindent
For metrology and survey: C. Boudineau$^C$, R. Kus$^D$, 
F.~Loeffler$^D$ and the DESY survey team. 

\noindent
For special monitor information from HERA and the design of 
the interlock system: P.~Duval$^D$, S.~Herb$^D$,
K.-H.~Mess$^D$, F. Peters$^D$, W. Schuette$^D$, M. Wendt$^D$.

\noindent
For installation and services: W. Beckhusen$^D$, H. Grabe-Celik$^D$, 
G. Kessler$^D$,  
G. Meyer$^D$, N.~Meyners$^D$, W. Radloff$^D$, U. Riemer$^D$
and F.R. Ullrich$^D$.    

\noindent
For developing elliptical cutting of detectors: N. Mezin$^C$ and I. Sexton$^C$.

\noindent
For vacuum design, testing and urgent repairs: R. Hensler$^D$, 
J. Kouptsidis$^D$, J. Roemer$^D$ and H-P. Wedekind$^D$.

\noindent
For front-end electronics design and front-end assembly: D.~Dorfan$^S$, 
J. De Witt$^S$, W.A. Rowe$^S$, E. Spencer$^S$ and A. Webster$^S$. 

\noindent
For the development of the special multi-layer boards which support the 
detectors and the front-end electronics: A. Gandi$^C$, 
L. Mastrostefano$^C$, C.~Millerin$^C$, A. Monfort$^C$, M. Sanchez$^C$ and 
D.~Pitzl$^S$, a former member of ZEUS.

\noindent
For the loan of S5 and S6 and part of their modification as well as help
with urgent repairs: B.~Jeanneret$^C$, 
R.~Jung$^C$, R.~Maleyran$^C$ and M.~Sillanoli$^C$. 

\noindent
For service, control and readout electronics:
F. Benotto$^T$,   M.~Ferrari$^B$, H.~Larsen$^L$, F. Pellegrino$^{Cs}$, 
J. Schipper$^L$, 
P.P. Trapani$^T$ and A. Zampieri$^T$.

\begin{thebibliography}{99}

\bibitem{hera_diffractive}
ZEUS Collab., M. Derrick et al., Phys. Lett. {\bf B 315} (1993) 481;\\
H1 Collab., T. Ahmed et al., Nucl. Phys. {\bf B429} (1994) 477;\\
ZEUS Collab., M. Derrick et al., Phys. Lett. {\bf B 332} (1994) 228;\\
H1 Collab., T. Ahmed et al., Phys. Lett. {\bf B 348} (1995) 681;\\
ZEUS Collab., M. Derrick et al., Z. Phys. {\bf C 68} (1995) 569;\\
ZEUS Collab., M. Derrick et al., Phys. Lett. {\bf B 346} (1995) 399;\\
ZEUS Collab., M. Derrick et al., Phys. Lett. {\bf B 356} (1995) 129;\\ 
ZEUS Collab., M. Derrick et al., Z. Phys. {\bf C 70} (1996) 391.

\bibitem{rho93} ZEUS Collab., M. Derrick et al., Z. Phys. {\bf C 69} (1995) 39. 
\bibitem{rhoh1}H1 Collab., S. Aid et al., Nucl. Phys. {\bf B463} (1996) 3.
\bibitem{h1psi} H1 Collab., T. Ahmed et al., Phys. Lett. {\bf B 338} (1994) 507;\\
H1 Collab., S. Aid et al., DESY report DESY 96-037 (1996).
\bibitem{psizeus} ZEUS Collab., M. Derrick et al., Phys. Lett. {\bf B 350} (1995) 120.
\bibitem{rhodis93}ZEUS Collab., M. Derrick et al., Phys. Lett. {\bf B 356} (1995) 601.
\bibitem{H1_rhopsi_hiq2}H1 Collab., S. Aid et al., 
Nucl. Phys. {\bf B468} (1996) 3.

\bibitem{romanpot} U. Amaldi et al., Phys. Lett. {\bf 43B} (1973) 231.

\bibitem{bauer} For a review, see e.g. 
T.H. Bauer et al., Rev. Mod. Phys. {\bf 50} (1978) 261.
\bibitem{egloff} R. M. Egloff et al., Phys. Rev. Lett. {\bf 43} (1979) 657.
\bibitem{omega} OMEGA Photon Collab., D. Aston et al., Nucl. Phys. {\bf B209} (1982) 56.


\bibitem{detector_a}
ZEUS Collab., M.~Derrick et al., The ZEUS Detector, Status Report 1993, 
DESY (1993).
\bibitem{detector_b}
ZEUS Collab., M.~Derrick et al., Phys.~Lett. {\bf B 293} (1992) 465.


\bibitem{vxd}
C.~Alvisi et~al.,
Nucl.~Instr. and Meth. {\bf A305} (1991) 30.

\bibitem{ctd} N. Harnew et al., Nucl.~Instr.~Meth.~{\bf A279} (1989) 290;\\
B.Foster et al., Nucl.~Phys.,~Proc.~Suppl.~{\bf B32} (1993) 181;\\
B.Foster et al., Nucl.~Instr.~Meth.~{\bf A338} (1994) 254.
                            


\bibitem{CAL} M.Derrick et al., Nucl.~Instr.~Meth.~{\bf A309} (1991) 77;\\
A.Andresen et al., Nucl.~Instr.~Meth.~{\bf A309} (1991) 101;\\
A.Bernstein et al., Nucl.~Instr.~Meth.~{\bf A336} (1993) 23;\\
A.Caldwell et al., Nucl.~Instr.~Meth.~{\bf A321} (1992) 356.
                                          
\bibitem{lumi}
J.~Andruszk\'ow et al., DESY report DESY~92-066 (1992);\\
ZEUS Collab., M.Derrick et al., Z. Phys. {\bf C 63} (1994) 391.

\bibitem{ua4} R. Battiston et al., Nucl. Instr. and Meth. {\bf A238} (1985) 35.

  

\bibitem{tekz} D.E. Dorfan, Nucl. Instr. and Meth. {\bf A342} (1994) 143;\\
E. Barberis et al., Nucl. Phys. {\bf B32} (Proc. Suppl.) (1993) 513.
\bibitem{dtsc} J. DeWitt, Nucl. Instr. and Meth. {\bf A288} (1990) 209;\\
J. DeWitt, ``The Time Slice Chip", Senior Thesis (1989), UC Santa Cruz, 
Santa Cruz Institute for Particle Physics report SCIPP 89-24.

\bibitem{FNC} 
ZEUS Collab., ``A Forward Neutron Calorimeter for ZEUS", DESY PRC 93-08;\\
S. Bhadra et al., Nucl. Instr. and Meth. {\bf A354} (1995) 479. 

\bibitem{tesi_roberto}
R. Sacchi, ``Studio di eventi diffrattivi con uno spettrometro per
protoni a ZEUS", Tesi di Dottorato, University of Torino (1996), unpublished (in Italian).
\bibitem{dipsi} M. Arneodo, L. Lamberti and M. G. Ryskin,
DESY Report DESY 96-149 (1996), to appear in
Comp. Phys. Comm.

\bibitem{shilling-wolf} K. Schilling et al., Nucl. Phys. {\bf B15} (1970) 
397;\\
K. Schilling and G. Wolf, Nucl. Phys. {\bf B61} (1973) 381.         

\bibitem{pythia}
T. Sj\"{o}strand, Comp. Phys. Comm. {\bf 82} (1994) 74.

\bibitem{chapin}
T.J. Chapin et al., Phys.~Rev. {\bf D31} (1985) 17.

\bibitem{CDF} CDF Collab., F. Abe et al., Phys. Rev. {\bf D50} (1994) 5535.

\bibitem{kurek} K. Kurek, private communication.

\bibitem{drell} S.D. Drell, Phys. Rev. Lett. {\bf 5} (1960) 278.

\bibitem{soeding}
P. S\"{o}ding, Phys. Lett. {\bf 19} (1966) 702.

\bibitem{laa} G. Anzivino et al., Rivista del Nuovo Cimento, Vol. 13, no. 5 
(1990) 1.

\end{thebibliography}
\end{document}